\RequirePackage{fix-cm}
\documentclass[twoside,twocolumn,9pt]{article}
\usepackage{extsizes}
\usepackage{tabularx} 
\usepackage[super,sort&compress,comma]{natbib} 
\usepackage[version=3]{mhchem}
\usepackage[left=1.5cm, right=1.5cm, top=1.785cm, bottom=2.0cm]{geometry}
\usepackage{balance}
\usepackage{titletoc}
\usepackage{mathptmx}
\usepackage{sectsty}
\usepackage{graphicx} 
\usepackage{lastpage}
\usepackage{amsmath}
\usepackage{graphicx}
\usepackage{float}
\usepackage{braket}
\usepackage{siunitx}
\usepackage{multicol}
\usepackage{parskip}
\usepackage[format=plain,justification=justified,singlelinecheck=false,font={stretch=1.125,small,sf},labelfont=bf,labelsep=space]{caption}
\usepackage{float}
\usepackage{fnpos}
\usepackage[english]{babel}
\addto{\captionsenglish}{%
  
}
\usepackage{array}
\usepackage{droidsans}
\usepackage{charter}
\usepackage[T1]{fontenc}
\usepackage[usenames,dvipsnames]{xcolor}
\usepackage{setspace}
\usepackage[compact]{titlesec}
\usepackage{microtype}
\usepackage{hyperref}

\usepackage{epstopdf}

\definecolor{cream}{RGB}{222,217,201}
\begin{document}

\makeFNbottom
\makeatletter
\renewcommand\LARGE{\@setfontsize\LARGE{15pt}{17}}
\renewcommand\Large{\@setfontsize\Large{12pt}{14}}
\renewcommand\large{\@setfontsize\large{10pt}{12}}
\renewcommand\footnotesize{\@setfontsize\footnotesize{7pt}{10}}
\makeatother

\renewcommand{\thefootnote}{\fnsymbol{footnote}}
\renewcommand\footnoterule{\vspace*{1pt}%
\color{cream}\hrule width 3.5in height 0.4pt \color{black}\vspace*{5pt}} 
\setcounter{secnumdepth}{5}

\makeatletter 
\renewcommand\@biblabel[1]{#1}            
\renewcommand\@makefntext[1]%
{\noindent\makebox[0pt][r]{\@thefnmark\,}#1}
\makeatother 
\renewcommand{\figurename}{\small{Fig.}~}
\sectionfont{\sffamily\Large}
\subsectionfont{\normalsize}
\subsubsectionfont{\bf}
\setstretch{1.125} 
\setlength{\skip\footins}{0.8cm}
\setlength{\footnotesep}{0.25cm}
\setlength{\jot}{10pt}
\titlespacing*{\section}{0pt}{4pt}{4pt}
\titlespacing*{\subsection}{0pt}{15pt}{1pt}

\makeatletter 
\newlength{\figrulesep} 
\setlength{\figrulesep}{0.5\textfloatsep} 

\newcommand{\topfigrule}{\vspace*{-1pt}%
\noindent{\color{cream}\rule[-\figrulesep]{\columnwidth}{1.5pt}} }

\newcommand{\botfigrule}{\vspace*{-2pt}%
\noindent{\color{cream}\rule[\figrulesep]{\columnwidth}{1.5pt}} }

\newcommand{\dblfigrule}{\vspace*{-1pt}%
\noindent{\color{cream}\rule[-\figrulesep]{\textwidth}{1.5pt}} }

\makeatother

\twocolumn[
\begin{@twocolumnfalse}

\noindent
{\color{cream}\rule{\textwidth}{2pt}}

\vspace{0.6cm}

\begin{center}
\sffamily



{\LARGE\bfseries
 Long lived localized defect states in monolayer WSe$_2$: Optical Lifetime distribution and thermal evolution
\par}

\vspace{0.45cm}

{\large
Immanuel Thekkooden\textsuperscript{a,e,f},
Susmita Jana\textsuperscript{b,c},
Mrinal Deka\textsuperscript{d},
B.~R.~K.~Nanda\textsuperscript{b,c},
and V.~Praveen Bhallamudi\textsuperscript{a,e,f}
\par}

\vspace{0.45cm}

{\footnotesize
\textsuperscript{a}Department of Electrical Engineering,
Indian Institute of Technology Madras, Chennai 600036, India
\par}

\vspace{0.08cm}

{\footnotesize
\textsuperscript{b}Condensed Matter Theory and Computational Lab,
Department of Physics, Indian Institute of Technology Madras,
Chennai 600036, India
\par}

\vspace{0.08cm}

{\footnotesize
\textsuperscript{c}Center of Atomistic Modelling and Materials Design,
Indian Institute of Technology Madras, Chennai 600036, India
\par}

\vspace{0.08cm}

{\footnotesize
\textsuperscript{d} Center for 2D Materials Research and Innovation,  Department of Physics, Indian Institute of Technology, Madras. Chennai 600026, India
\par}

\vspace{0.08cm}

{\footnotesize
\textsuperscript{e}Quantum Defects Lab, Department of Physics,
Indian Institute of Technology Madras, Chennai 600036, India
\par}

{\footnotesize
\textsuperscript{f} Quantum Center of Excellence for Diamond and Emerging Materials (QuCenDiEM) Group,  Department of Physics, Indian Institute of Technology, Madras. Chennai 600026, India
\par}

\vspace{0.25cm}

{\footnotesize
\textsuperscript{*}Corresponding author:
\href{mailto:thekkoodenimmanuel2@gmail.com}
{\nolinkurl{praveen.bhallamudi@iitm.ac.in }}
\par}

\end{center}

\vspace{0.45cm}

\noindent
{\sffamily\normalsize\bfseries Abstract}

\vspace{0.15cm}

\noindent
{\normalsize
We carefully investigate the recombination dynamics of localized defect emission
in monolayer WSe$_2$  on SiO$_2$/Si substrate using time-resolved photoluminescence over the
temperature range of 4 K to 120 K. We observe  two long-lived optical lifetimes, one few nanosecond and one hundreds of nanoseconds. PL decay profile of these long lived states is fit well by a power-law, and based on laser fluence studies, the likely origin of the power-law is due to a distribution of  life-times rather than many body interactions. Temperature-dependence of these
rates shows that thermal detrapping 
governs these long-lived channels and we obtained a value of  60 meV for the characteristics energy in a Bose-Einstein  model. We performed spin-resolved density functional calculations
for selenium vacancies, the most likely source of native defects, to elucidate on spin- and momentum-forbidden pathways which are the  likely origin of these long lifetimes. Such detailed understanding of long-lived defect states is crucial for quantum and optoelectronic applications.
\par}

\vspace{0.55cm}

\noindent
{\color{cream}\rule{\textwidth}{2pt}}

\vspace{0.7cm}

\end{@twocolumnfalse}
]

\renewcommand*\rmdefault{bch}\normalfont\upshape
\rmfamily
\section*{}
\vspace{-1cm}

%

\section{Introduction}
Defects are ubiquitous in semiconductors and can profoundly alter their
electronic and optical properties by introducing localized states within
the band gap. These states provide carrier-capture and recombination
pathways that differ fundamentally from those of delocalized band-edge
excitons\cite{shin2020ultrafast,carbone2025creation} and can be important for applications including optical
memories, carrier-storage devices, and quantum-photonic technologies\cite{immanuel2022quantum,parto2021defect,li2019defect}.

Localized defect states in monolayer WSe$_2$ are particularly
interesting because they can exhibit sub-microsecond recombination
dynamics\cite{moody2018microsecond,wagner2021trap,qian2020defect,LiQuantum2022}, much longer than the typical picosecond-scale recombination of
free excitons\cite{godde2016exciton,robert2017fine,lopion2020temperature,huang2016probing}. WSe$_2$ has become one of the most widely studied TMDC
platforms for localized quantum emitters, a tendency that has been
proposed to arise partly from its dark-excitonic ground-state
configuration and strong susceptibility to exciton localization. These
characteristics make defect-localized states in WSe$_2$ promising for
quantum-photonic and optoelectronic applications\cite{ayari2018radiative}. However, only a
limited number of studies have reported microsecond-scale recombination
dynamics in WSe$_2$, and more detailed experiments are needed to better
understand the microscopic origin of these long-lived states and,
phenomenologically, the effects of the local charge and strain
environments on their dynamics\cite{rhodes2019disorder}.

In this work, we analyse defect-related emission from an exfoliated
monolayer of WSe$_2$ on a SiO$_2$/Si substrate in a more comprehensive
manner than previously reported. We first identify the defect emission
through the temperature and laser-power dependence of the different
emission peaks. We then use a band-pass filter to isolate the
defect-related emission and perform time-resolved measurements of the
optical recombination dynamics over tens of microseconds. We find that
the decay contains a dominant sub-nanosecond exponential component
together with two weak, long-lived decay channels. A power-law model
best describes the decay profile of these long-lived channels.

Combined
with their weak excitation-fluence dependence, this observation supports
an interpretation based on a broad distribution of defect-mediated
recombination times resulting from inhomogeneity in the local,
likely charge and strain, environment. Our temperature-dependent
measurements further reveal phonon-assisted detrapping as a primary relaxation mechanism. A
Bose--Einstein model describes the corresponding rate variation and
yields a characteristic energy of approximately
60~meV. Finally, spin-resolved density-functional-theory and
optical-transition calculations identify localized in-gap states and
additional weak momentum- and spin-relaxed recombination pathways,
providing a plausible microscopic basis for the observed long-lived
emission.

\section{Results}

\begin{figure*}
 \centering
 \includegraphics[height=10 cm]{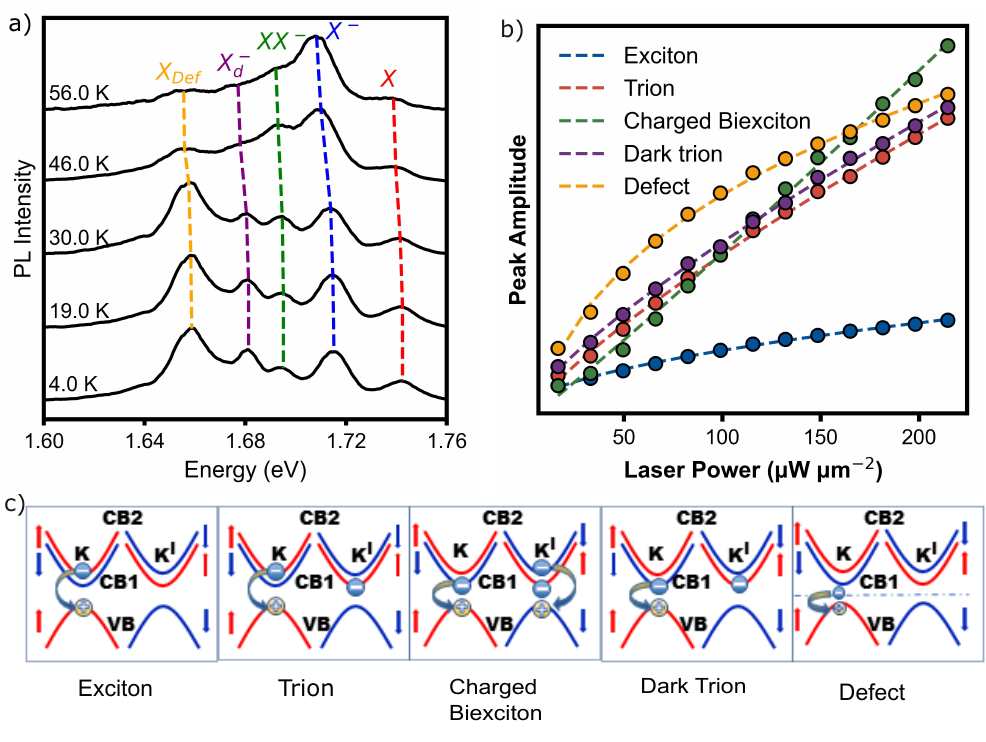}
 \caption{PL studies of monolayer WSe$_2$ as function of temperature and laser power. a) PL peaks at 4 K shows distinct five peak corresponding to different quasiparticle. b) PL peak position and behaviour of each peak under different laser power are used to identify quasi-particle. The dots represent data point and solid line indicates fit. Defect peak shows fast saturation with laser intensity. c) Spin valley configuration of each quasi-particle. Ground state of monolayer WSe$_2$ is dark in nature. }
 \label{fgr:example2col}
\end{figure*}

The monolayer WSe$_2$ sample was obtained from exfoliating bulk crystal using scotch-tape method. Monolayer region was first identified from optical contrast under a optical microscope. To confirm the monolayer, we did room temperature (RT) PL measurements and Raman measurements (see ESI). PL spectrum reveals a peak at 1.65 eV which is typical for monolayer sample\cite{sahin2013anomalous}. PL intensity was very high indicating  the  presence of direct bandgap which is only present in monolayer form. Had it been bi-layer or higher number of layers, PL peak energy would have been very weak and redshifted\cite{huang2016probing,sahin2013anomalous} at RT .  The Raman spectra (shown in ESI) of the sample was taken at RT using 532 nm laser. The Raman spectra showed characteristics two modes, one mode at \qty{246.5}{\per\centi\meter} and  other mode at \qty{257.2}{\per\centi\meter}, with a distance between them as 10.61  $cm^{-1}$, which is typical for monolayer samples\cite{tonndorf2013photoluminescence,sahin2013anomalous,wagner2021trap}. Also absence of breathing B peak at 308 $cm^{-1}$ confirms that this is monolayer sample\cite{lopion2020temperature}. Thus from RT PL and Raman studies , we conclude that sample is monolayer.

Sample was cooled and PL measurements were taken as shown in Figure \ref{fgr:example2col} a.

\subsection{Assignment of PL peaks in monolayer \texorpdfstring{WSe$_2$}{WSe2}}
At room temperature, the PL spectrum of monolayer WSe$_2$ is dominated by the neutral exciton ($X$), while a weak low-energy shoulder appears approximately 27 meV below $X$ (see ESI). This energy separation is consistent with previous reports of the bright negatively charged intervalley trion ($X^{-}$), comprising an optically bright electron--hole pair in one valley and an additional electron in the opposite valley as shown in \ref{fgr:example2col} c . As the temperature is lowered, the $X^{-}$ emission increases in intensity and becomes clearly distinguishable from $X$ below approximately 120 K. Both the $X$ and $X^{-}$ peaks redshift with increasing temperature, consistent with band-gap renormalization arising from lattice expansion and enhanced electron--phonon interactions as shown in ESI \cite{davila2024temperature}.

\begin{table*}[t]
\small
\centering
\caption{Tentative assignment of the low-temperature PL peaks based on their measured energy separation from the neutral exciton and excitation-power dependence.}
\label{tab:pl_peak_identification}
\begin{tabular*}{\textwidth}{@{\extracolsep{\fill}}llll}
\hline
Quasiparticle &
Measured energy separation from $X$ (meV) &
Reported separation (meV) &
Measured power exponent $\alpha$ \\
\hline
Neutral exciton ($X$) &
0 &
-- &
$0.57 \pm 0.04$ \\

Bright intervalley trion ($X^-$) &
27.4 &
28\cite{li2018revealing} &
$0.79 \pm 0.03$ \\

Negatively charged biexciton ($XX^-$) &
47.5 &
49\cite{li2018revealing}, 52\cite{liu2019gate} &
$1.06 \pm 0.05$ \\

Dark trion ($X_d^-$) &
61.4 &
57\cite{liu2019gate} &
$0.77 \pm 0.02$ \\

Localized defect-related emission ($X_{\mathrm{Def}}$) &
87.0 &
92\cite{godde2016exciton,you2015observation},
\cite{wang2014valley,huang2016probing} &
$0.33 \pm 0.04$ \\
\hline
\end{tabular*}
\end{table*}

Below approximately 65 K, five distinct emission features are resolved in the PL spectrum. Two are assigned to $X$ and $X^{-}$, while three additional low-energy peaks become progressively more pronounced as the temperature approaches 4 K. To examine their origin, excitation-power-dependent PL measurements were performed as shown in Fig \ref{fgr:example2col} b. At each laser power, the complete spectrum was fitted using a sum of five Lorentzian functions, and the integrated area of each Lorentzian component was extracted. The resulting integrated intensities were fitted using the power-law relation
\(I \propto P^{\alpha}\), where \(P\) is the excitation-power density and
\(\alpha\) is the power-law exponent. The measured values of $\alpha$, together with the energy separation of each peak from the neutral exciton and its temperature evolution, were used collectively to assign the observed emission features\cite{huang2016probing}. The resulting assignments are summarized in Table~\ref{tab:pl_peak_identification}.

In the low-excitation limit, exciton and trion emission is generally expected to exhibit an approximately linear power dependence, $\alpha\approx1$, whereas biexcitonic complexes are expected to exhibit a superlinear dependence that approaches $\alpha\approx2$\cite{huang2016probing,ye2018efficient}. Experimentally measured exponents, however, often deviate from these ideal values because the quasiparticle populations are not necessarily in thermodynamic equilibrium and are influenced by carrier capture, non-radiative relaxation, state filling, and many-body interactions\cite{huang2016probing,Mouri_2014}. In the present measurements, the neutral-exciton emission exhibits a sublinear dependence, with $\alpha_X=0.57\pm0.04$. This behavior may arise from a combination of defect-assisted exciton capture, hot-carrier generation under non-resonant excitation, saturation of available radiative channels, and exciton--exciton annihilation at elevated carrier densities\cite{you2015observation,phillips1992biexciton,kim1994thermodynamics,Mouri_2014}. Consequently, the measured power exponents are not treated as independent fingerprints but are interpreted together with the spectral positions and temperature dependence of the corresponding peaks.

Spin--orbit coupling splits both the valence and conduction bands near the K and K$^\prime$ valleys of monolayer WSe$_2$. The valence-band splitting is substantially larger than the conduction-band splitting\cite{kosmider2013large,ren2023measurement,le2015spin}. The relative spin ordering of the upper valence band and lowest conduction band gives monolayer WSe$_2$ a dark excitonic ground state, while transitions involving the higher conduction band form the bright excitonic manifold\cite{kapuscinski2021rydberg}. Spin-dark excitons possess predominantly out-of-plane transition dipoles and therefore emit only weakly along the surface normal. Their emission can be enhanced or collected using in-plane excitation geometries\cite{wang2017plane,tang2019long}, plasmonic coupling\cite{zhou2017probing,lo2022plasmonic,tang2019long}, or high-numerical-aperture objectives that collect light emitted at large angles\cite{robert2017fine,li2018revealing}.

The high-NA objective employed in the present experiment therefore permits partial collection of emission from dark excitonic species. The peak located 61.4 meV below $X$ exhibits a sublinear power dependence of $\alpha=0.77\pm0.02$ and decreases in intensity with increasing temperature. Its energy separation and thermal behavior are consistent with a tentative assignment to the negatively charged dark trion, $X_d^{-}$. Within the commonly used intervalley picture, this complex contains a dark electron--hole pair associated with one valley and an additional electron in the opposite valley\cite{liu2019gate}. Thermal redistribution of carriers from the lower dark manifold to the higher bright manifold reduces the population of dark excitonic complexes with increasing temperature, while increasing the relative contribution of optically bright states. 

The assignment of the peak located 47.5 meV below $X$ requires particular care because its spectral position lies close to the reported energies of both neutral dark exciton and charged biexcitonic complexes. A negatively charged biexciton, $XX^{-}$, can be described as a correlated complex involving bright and dark exciton constituents together with an additional resident electron\cite{ye2018efficient,barbone2018charge}. The measured exponent of this peak is $\alpha_{XX^-}=1.06\pm0.05$, which is approximately twice the measured exponent of the neutral exciton, $\alpha_{XX^-}\approx2\alpha_X$.

This relative scaling, together with the measured energy separation, supports a tentative assignment of the peak to $XX^{-}$. The decrease of this emission with increasing temperature is also consistent with thermally activated dissociation of biexcitonic complexes\cite{ye2018efficient,zhang2015experimental,kipczak2024impact}. However, definitive discrimination between $XX^{-}$ and nearby neutral dark-excitoni feature would require additional magneto-optical, polarization-resolved, or gate-dependent measurements. The $XX^{-}$ label is therefore used here as the assignment most consistent with the available experimental evidence rather than as an unambiguous microscopic identification.

The lowest-energy peak, located approximately 87 meV below $X$, exhibits the strongest sublinear power dependence, with $\alpha=0.33\pm0.04$, and approaches saturation rapidly as the excitation power is increased. Such behavior is characteristic of emission from a finite population of localized excitonic states\cite{tongay2013defects,huang2016probing}. Atomic defects can introduce localized electronic states within the band gap, which capture electrons or holes and subsequently participate in defect-bound exciton recombination\cite{carbone2025creation}. 

The large energy separation from the neutral exciton, pronounced low-temperature emission, strongly sublinear power dependence, and rapid thermal quenching collectively support assignment of this feature to localized defect-related emission, denoted $X_{\mathrm{Def}}$. Here, $X_{\mathrm{Def}}$ is used as a phenomenological spectral label and does not, by itself, specify the precise atomic configuration responsible for the emission. The suppression of this peak at elevated temperatures is consistent with thermally activated escape from localized states and the opening of competing non-radiative recombination pathways\cite{parto2021defect}. The possible microscopic origin of these localized states, including the relative roles of intrinsic point defects and local strain, is discussed separately using the sample preparation, defect-formation energetics, and electronic-structure calculations.

Overall, the assignments of $X$, $X^{-}$, $XX^{-}$, $X_d^{-}$, and $X_{\mathrm{Def}}$ are based on the combined agreement between the measured energy separations, excitation-power exponents, and temperature-dependent PL behavior. While the $X$,$X^{-}$ and $X_{\mathrm{Def}}$ assignments are well established, the $XX^{-}$ and $X_d^{-}$ assignments should be regarded as tentative in the absence of gate-dependent or magneto-optical measurements.

\subsection{Long Lifetime from Defect Emission}

\begin{figure*}
 \centering
 \includegraphics[height=9.8 cm]{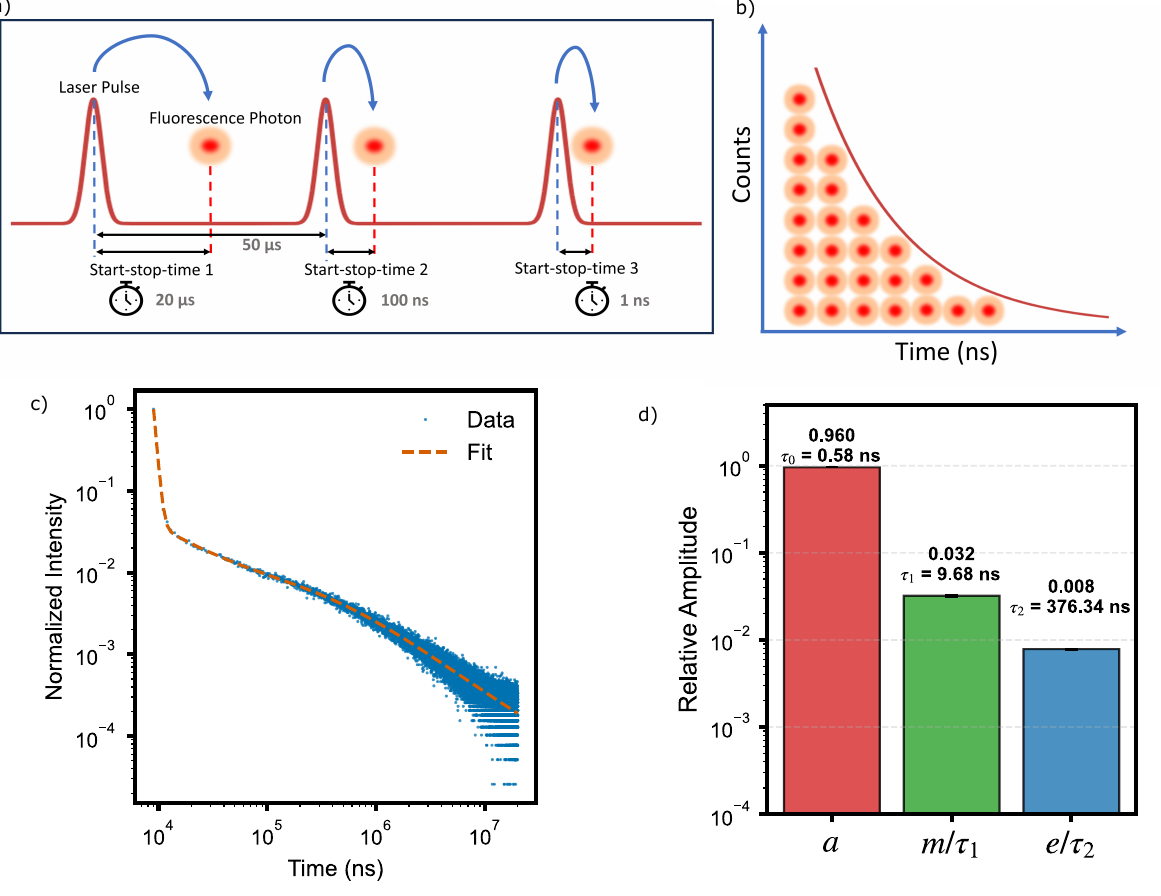}
 \caption{Lifetime analysis of the defect-related emission.
(a) Schematic of the time-correlated single-photon counting (TCSPC)
measurement, in which each excitation pulse provides the START signal
and the detection of an emitted photon provides the STOP signal.
(b) Schematic of the photon-arrival-time histogram accumulated over
repeated excitation cycles.
(c) Defect-selective PL decay measured using a 750~nm band-pass filter
and fitted with the model described in the text. The discrete count
levels observed at late times arise from the low photon-count rate and
detector dark counts.
(d) Relative amplitude contributions and characteristic timescales of
the individual recombination channels extracted from the fit.}
 \label{longlifetime}
\end{figure*}

Time-resolved photoluminescence (TRPL) measurements were performed for each quasiparticle using appropriate band-pass filters (see ESI). For the defect emission, a 750 nm band-pass filter with a full-width at half-maximum (FWHM) of 10 nm was employed to selectively isolate the defect-related PL. Following nonresonant excitation at 525 nm, photoexcited electron go up in conduction band and then undergo ultrafast relaxation toward the band-edge excitonic
manifold\cite{steinleitner2017direct}. A fraction of the excitonic
population may subsequently be captured by defect-related localized
potentials, giving rise to localized exciton
\cite{steinleitner2017direct,liu2019neutral,shin2020ultrafast,li2018auger}. Radiative recombination of these defect-bound excitons gives rise to the defect emission observed at 747 nm. The photon arrival histogram was acquired over a temporal window of 50 $\mu$s using the TCSPC scheme illustrated in Fig.~\ref{longlifetime}(a,b), ensuring that the complete decay dynamics of the defect emission were captured. 

The measured decay as shown in  Fig.~\ref{longlifetime}(c) could not be satisfactorily described using conventional mono-, bi-, tri or four exponential functions (see ESI). Although an empirical fit using five exponential components reproduced the data, such a model introduces a large number of unconstrained fitting parameters without providing a physically meaningful interpretation of the underlying recombination processes. We therefore adopted a phenomenological model consisting of one exponential decay channel together with two power-law components,

\begin{equation}
y(t)=
a\exp\left(-\frac{t}{\tau_{0}}\right)
+
\frac{m}{t+\tau_{1}}
+
\frac{e}{t+\tau_{2}}
+
\mathrm{baseline},
\label{decaymodel}
\end{equation}

where the exponential term represents the dominant fast recombination channel, while the two inverse-time components account for the experimentally observed long-lived non-exponential dynamics.

The extracted parameters as shown in Fig.~\ref{longlifetime}(d) reveal a fast recombination component with a characteristic lifetime of approximately 500 ps together with two long-lived channels characterized by $\tau_{1}\approx9$ ns and $\tau_{2}\approx376$ ns. The corresponding amplitude contributions are approximately 96\%, 3.2\%, and 0.7\% for the $\tau_{0}$, $\tau_{1}$, and $\tau_{2}$ channels, respectively, as summarized in Fig.~\ref{longlifetime}(d). The dominant sub-nanosecond channel is attributed to the primary radiative recombination pathway of defect-bound excitons. In contrast, the two weaker long-lived channels are considerably slower than the radiative lifetime of free excitons and are therefore expected to originate from weakly allowed defect-assisted recombination processes\cite{dass2019ultra}.

A non-exponential $\frac{1}{t}$ decay has previously been employed to describe the dynamics of free excitons in monolayer WSe$_2$ in the context of exciton-exciton annihilation (EEA)\cite{xu2024control,erkensten2021dark,liu2019neutral}. Similar power-law relaxation has also been reported in a variety of semiconductor systems, including shallow defects, quantum dots, color centers and perovskites in long time scale measurement\cite{yuan2024shallow,andersen2003temperature,PhysRevA.77.042719,kuno2000nonexponential,houel2015autocorrelation,akmaev2020nonexponential}. To the best of our knowledge, however, such power-law behaviour has not previously been reported for localized defect emission in monolayer TMDCs.

 We attribute the observed power-law behaviour to a continuous distribution of defect recombination rates arising from the inhomogeneous electrostatic environment experienced by localized defect states. Unlike free excitons, defect-bound excitons remain spatially localized and therefore probe only their immediate dielectric surroundings. Several factors can contribute to such an inhomogeneous potential landscape, including nanoscale topographical corrugations of the underlying amorphous SiO$_2$ substrate\cite{chaudhary2020origin,hernandez2022strain} and local charge puddles arising from dangling bonds and surface adsorbates\cite{martin2008observation,zhang2010origin,rhodes2019disorder}. Consequently, defects located at different spatial positions experience different local dielectric environments, giving rise to a distribution of recombination rates rather than a single characteristic lifetime.

The above interpretation can be formulated quantitatively by expressing each power-law component as a superposition of exponentially decaying channels with a continuous distribution of recombination rates. For a power-law component characterized by a timescale $\tau_i$,

\begin{equation}
I_i(t)=\frac{C_i}{t+\tau_i},
\qquad i=1,2,
\label{eq:powerlaw_component}
\end{equation}

the initial intensity is $I_i(0)=C_i/\tau_i$. The normalized decay can therefore be written as

\begin{equation}
\frac{I_i(t)}{I_i(0)}
=
\frac{\tau_i}{t+\tau_i}
=
\int_{0}^{\infty}
P_{\Gamma,i}(\Gamma)
e^{-\Gamma t}\,d\Gamma,
\label{eq:rate_superposition}
\end{equation}

where $P_{\Gamma,i}(\Gamma)$ is the normalized probability density of local recombination rates and $\Gamma=1/\tau$ is the recombination rate associated with an individual localized state. Using the Laplace-transform identity

\begin{equation}
\frac{\tau_i}{t+\tau_i}
=
\int_{0}^{\infty}
\tau_i e^{-\Gamma\tau_i}
e^{-\Gamma t}\,d\Gamma,
\end{equation}

the corresponding rate distribution is obtained as

\begin{equation}
P_{\Gamma,i}(\Gamma)
=
\frac{1}{\Gamma_i}
\exp\left(-\frac{\Gamma}{\Gamma_i}\right),
\qquad
\Gamma_i=\frac{1}{\tau_i},
\label{eq:rate_distribution}
\end{equation}

with

\begin{equation}
\int_{0}^{\infty}
P_{\Gamma,i}(\Gamma)\,d\Gamma=1.
\end{equation}

The corresponding lifetime distribution is obtained using the transformation

\begin{equation}
P_{\tau,i}(\tau)
=
P_{\Gamma,i}(\Gamma)
\left|
\frac{d\Gamma}{d\tau}
\right|,
\qquad
\Gamma=\frac{1}{\tau}.
\end{equation}

Since $\left|d\Gamma/d\tau\right|=1/\tau^{2}$, the normalized lifetime distribution becomes

\begin{equation}
P_{\tau,i}(\tau)
=
\frac{\tau_i}{\tau^{2}}
\exp\left(-\frac{\tau_i}{\tau}\right),
\qquad \tau>0.
\label{eq:lifetime_distribution}
\end{equation}



The most probable lifetime is obtained by maximizing Eq.~(\ref{eq:lifetime_distribution}),

\begin{equation}
\frac{dP_{\tau,i}(\tau)}{d\tau}=0,
\end{equation}

which yields

\begin{equation}
\tau_{\mathrm{max},i}
=
\frac{\tau_i}{2}.
\label{eq:most_probable_lifetime}
\end{equation}

Accordingly, the fitted quantities $\tau_1$ and $\tau_2$ should not be interpreted as single discrete lifetimes. Instead, they define the characteristic timescales of two distinct lifetime distributions. For the measured values $\tau_1\approx9$ ns and $\tau_2\approx376$ ns, the corresponding distributions attain their maxima near 4.5 ns and 188 ns, respectively. The presence of two well-separated distributions indicates that the long-lived defect emission contains two distinct classes of slow defect-assisted recombination dynamics, each spanning a broad range of local lifetimes. The amplitude-weighted forms of this distributions, are presented in the ESI.

The inverse-Laplace representation provides a direct physical interpretation of the measured power-law relaxation. Defects located at different spatial positions experience different local electrostatic and dielectric environments, resulting in site-dependent transition rates. The measured decay represents the ensemble average over these non-equivalent localized states. The observed non-exponential dynamics are therefore consistent with an inhomogeneous distribution of defect-bound exciton lifetimes rather than a small number of discrete defect levels.

Several independent checks were performed to establish that the extracted long-lived components do not originate from instrumental broadening or detector noise. The instrument response function (IRF) was measured using strongly attenuated laser reflection, with the optical power kept sufficiently low to minimize detector dead-time effects. The measured IRF was subsequently convolved numerically with the decay model in Eq.~(\ref{decaymodel}), and the model parameters were varied until the convolved decay reproduced the experimental trace. The parameters obtained from this convolution analysis were consistent with those extracted from direct fitting, as discussed in the ESI.

The fastest component, $\tau_0$, is comparable to the temporal width of
the instrument response function (IRF). Consequently, its extracted
value is subject to substantial instrumental uncertainty and is not
included in the detailed temperature- and laser-fluence-dependent
analysis. In contrast, $\tau_1$ and $\tau_2$ are substantially longer than the instrumental response and can be extracted reliably. Time-series and noise analyses further confirmed that the measurements were limited predominantly by photon shot noise. The discrete intensity levels visible at the end of the decay trace in Fig.~\ref{longlifetime}(c) arise from low photon count and quantized detector dark counts rather than from an additional physical decay channel. Details of the IRF measurement, convolution procedure, model comparison, and noise analysis are provided in the ESI.

Taken together, the defect-selective TRPL measurements establish the coexistence of a dominant fast exponential channel and two weak, broadly distributed long-lived recombination channels. The power-law components therefore reflect ensemble-averaged relaxation across non-equivalent localized environments rather than two sharply defined defect lifetimes. Having established the distributed nature of these channels, we next examine their fluence and temperature dependence to determine whether they arise from the same microscopic process or respond differently to thermal activation.

\subsection{Effect of fluence on life-time characteristics}

To investigate whether the decay process in our system is dominated by many-body interactions, we performed laser power-dependent lifetime as shown in ESI . Exciton-exciton annihilation (EEA) is a non-radiative many-body process in which two excitons interact such that one exciton recombines nonradiatively and transfers its energy to the other exciton, which is subsequently excited to a higher energy state\cite{li2018auger,deng2020long,shin2020ultrafast}. A key signature of many-body interactions such as EEA and defect assisted Auger recombination is a reduction in lifetime with increasing laser fluence, as non-radiative channels are activated in addition to the intrinsic radiative recombination, draining the exciton population more rapidly\cite{liu2019neutral,shin2020ultrafast,wang2015ultrafast}. In our study, we do not observe any reduction in lifetime with increasing laser power for $\tau_1$ and $\tau_2$. Therefore, we conclude that long-lived $\tau_1$ and $\tau_2$ channels are not significantly affected by many-body interactions such as EEA under the experimental conditions used\cite{LiQuantum2022}.

The excitation-fluence-dependent measurements were performed over an
incident fluence range of \(1.27\)--\(23.4~\mathrm{mJ\,cm^{-2}}\) per
pulse. As estimated in the ESI, this corresponds to an initial
free-exciton density of the order of \(10^{13}~\mathrm{cm^{-2}}\), a
regime in which exciton--exciton annihilation is generally expected to
be significant. Nevertheless, neither of the long-lived characteristic
times shows a systematic reduction with increasing fluence; instead,
their values fluctuate. This behaviour indicates that the long-lived defect-related
recombination is not dominated by exciton--exciton annihilation.
Spatial localization suppresses the diffusion-assisted encounters
required for efficient annihilation between defect-bound excitons.
Moreover, the long-lived emission persists on timescales at which the
free-exciton and trion populations have already decayed substantially (see ESI to see exciton and trion lifetime measurement),
thereby reducing their interaction with the recombination of localized defect
population. The combined effects of spatial localization and delayed
recombination therefore strongly suppress density-dependent many-body
interactions in the long-lived defect-emission channels. Thus effect of spatial disorder will be more on long-lifetimes of localised-defect exciton rather than many body interactions. Therefore, the observed $1/t$-like decay is more consistently attributed
to a disorder-induced distribution of recombination rates among
localized defect states than to density-dependent many-body interactions.

\subsection{Effect of temperature on life-time characteristics}

\begin{figure*}
  \centering
  \includegraphics[
        width=0.8\textwidth]{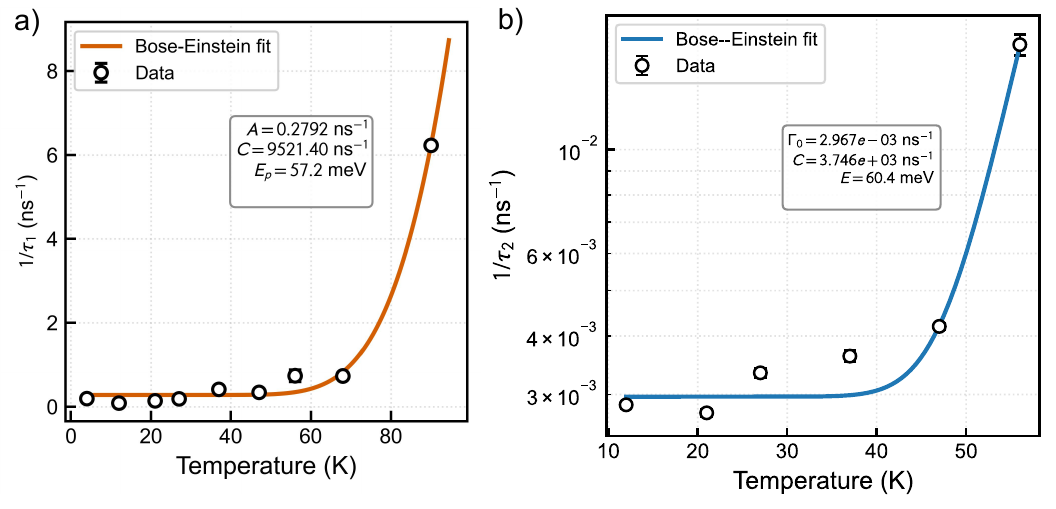}
   \caption{Temperature dependence of the defect-related recombination
dynamics. (a) The intermediate lifetime component, $\tau_{1}$, decreases
with increasing temperature. The corresponding recombination rate,
$1/\tau_{1}$, is fitted using a Bose--Einstein model. (b) The long-lived
component, $\tau_{2}$, shows a distinct temperature dependence, and its
recombination rate, $1/\tau_{2}$, is also described using a
Bose--Einstein model. This component can no longer be reliably resolved
above approximately 57~K.}
 \label{tempdependence}
 \end{figure*} 
. 

Temperature-dependent TRPL measurements were performed to investigate
the thermal evolution of the defect-bound exciton dynamics. Because the
defect-related PL intensity decreases with increasing temperature and
becomes negligible above approximately 120~K, the acquisition time was
increased at elevated temperatures to maintain a comparable
signal-to-noise ratio and reliably resolve the weak long-lived decay
components.

The temperature-dependent rates of both long-lived channels were
described using a Bose--Einstein phonon-occupation model,

\begin{equation}
\Gamma(T)
=
A+
\frac{C}
{\exp\!\left(\dfrac{E}{k_{\mathrm{B}}T}\right)-1},
\label{eq:bose_einstein_rate}
\end{equation}

where \(A\) represents the temperature-independent relaxation rate,
\(C\) parameterizes the strength of the phonon-assisted contribution,
and \(E\) is the corresponding characteristic phonon-related energy.
The intermediate component, \(\tau_{1}\), exhibits a pronounced
temperature dependence over the full measurement range and remains
resolvable up to approximately 120~K, as shown in
Fig.~\ref{tempdependence}(a). Its rate, \(1/\tau_{1}\), is well
described by Eq.~\ref{eq:bose_einstein_rate}, yielding a characteristic
energy of approximately 57~meV. This behaviour
indicates that the depopulation of the \(\tau_{1}\) channel is governed
predominantly by phonon-assisted non-radiative detrapping. It should be noted that the decay rate associated with $\tau_{1}$
increases with temperature and approaches the temporal resolution limit
of our measurement system. Therefore, the extracted value of
$57~\mathrm{meV}$ should be regarded as an upper bound on the
characteristic energy scale.


The longer-lived component, $\tau_{2}$, also exhibits a pronounced
temperature dependence, as shown in Fig.~\ref{tempdependence}(b). Its
recombination rate is well described by the same Bose--Einstein model,
yielding a characteristic energy of
$E = 60.43 \pm 2.23~\mathrm{meV}$. However, the amplitude associated
with this component decreases strongly with increasing temperature, and
$\tau_{2}$ can no longer be reliably resolved above approximately
57~K.
The phonon energy scale obtained for both long-lived chanels are closer to the combined energy of two optical phonons
in monolayer WSe$_2$ and is therefore consistent with a possible
two-phonon-assisted relaxation process\cite{wagner2021trap}.

Above approximately 120 K, the defect-related emission is completely quenched, leaving the PL spectrum dominated by bright exciton and trion emission. The overall observations therefore indicate that defect recombination in monolayer WSe$_2$ cannot be described by a single trapping pathway but instead involves multiple distributed relaxation channels with distinct microscopic origins.

\section{Discussions}

\subsection{Discussion on the microscopic origin of the defect states}

Several intrinsic defects have been theoretically predicted and experimentally observed in monolayer WSe$_2$, including selenium vacancies ($V_{\mathrm{Se}}$), tungsten vacancies ($V_{\mathrm{W}}$), antisite defects ($Se_{\mathrm{W}}$ and $W_{\mathrm{Se}}$), pore defects, and trefoil defects\cite{zhang2017defect,parto2021defect,lin2015three,lin2018realizing}. In the present work, the investigated flakes were mechanically exfoliated from high-quality flux-grown bulk crystals synthesized from atomistic precursors, and no intentional defect engineering, irradiation, plasma treatment, or annealing was performed after exfoliation. Consequently, the observed localized emission is expected to originate predominantly from native defects already present in the crystal.

Complex defect structures such as pore defects and trefoil defects are generally associated with CVD-grown samples or high-energy electron irradiation, where significant atomic reconstruction occurs\cite{qian2020defect,parto2021defect}. Likewise, antisite defects typically require atomic migration during high-temperature growth or post-growth thermal processing\cite{lin2018realizing}. Since none of these processes were involved in the preparation of the present samples, such defect configurations are expected to be less abundant than intrinsic point defects. Furthermore, tungsten vacancies possess substantially higher formation energies than selenium vacancies\cite{tosun2016air,nguyen2021gate}, making them considerably less probable under equilibrium growth conditions.

Among the intrinsic point defects, selenium vacancies have consistently been reported to possess the lowest formation energy\cite{zheng2019first,choi2023chemical,carin2024role}. The observed defect emission energy and unusually long radiative lifetime are also consistent with previous reports on localized excitons associated with selenium-vacancy-related defect states\cite{moody2018microsecond,qian2020defect}. In addition, the dominance of negatively charged excitonic complexes observed in the PL spectra suggests an n-type environment, which has frequently been attributed to donor-like selenium vacancies in WSe$_2$\cite{tosun2016air,zhang2017defect}. Although substrate-induced charge transfer may also contribute to the observed electron doping, the combined experimental observations are broadly consistent with selenium-vacancy-related defect states.

Recent studies have shown that oxygen can occupy selenium vacancy sites following exposure to ambient conditions, forming substitutional oxygen defects ($O_{\mathrm{Se}}$)\cite{barja2019identifying}. These defects have been proposed to passivate the deep gap states associated with bare selenium vacancies as shown in ESI, thereby suppressing carrier trapping and reducing localized emission\cite{barja2019identifying,zheng2019first,zheng2019point}. Since the present samples exhibit pronounced defect trapping manifested by long-lived PL decay, substitutional oxygen defects are unlikely to be the dominant origin of the observed localized states. Oxygen interstitials have also recently been proposed as potential sources of localized emitters in WSe$_2$, and therefore cannot be excluded\cite{zheng2019point}. Likewise, double selenium vacancies possess relatively low formation energies and remain possible candidates, although isolated selenium vacancies are generally expected to occur with higher probability\cite{carin2024role}.

We emphasize that the present measurements do not provide a definitive microscopic identification of the defect species. Such identification would require complementary atomic-resolution structural characterization, for example by scanning transmission electron microscopy or scanning probe techniques. Nevertheless, considering the sample preparation method, reported defect formation energetics, emission energy, long-lived trapping dynamics, and the observed n-type optical characteristics, the experimental evidence is most consistent with selenium-vacancy-related defect states.
               
\subsection{DFT calculations}
\begin{figure*}
 \centering
 \includegraphics[height=8cm]{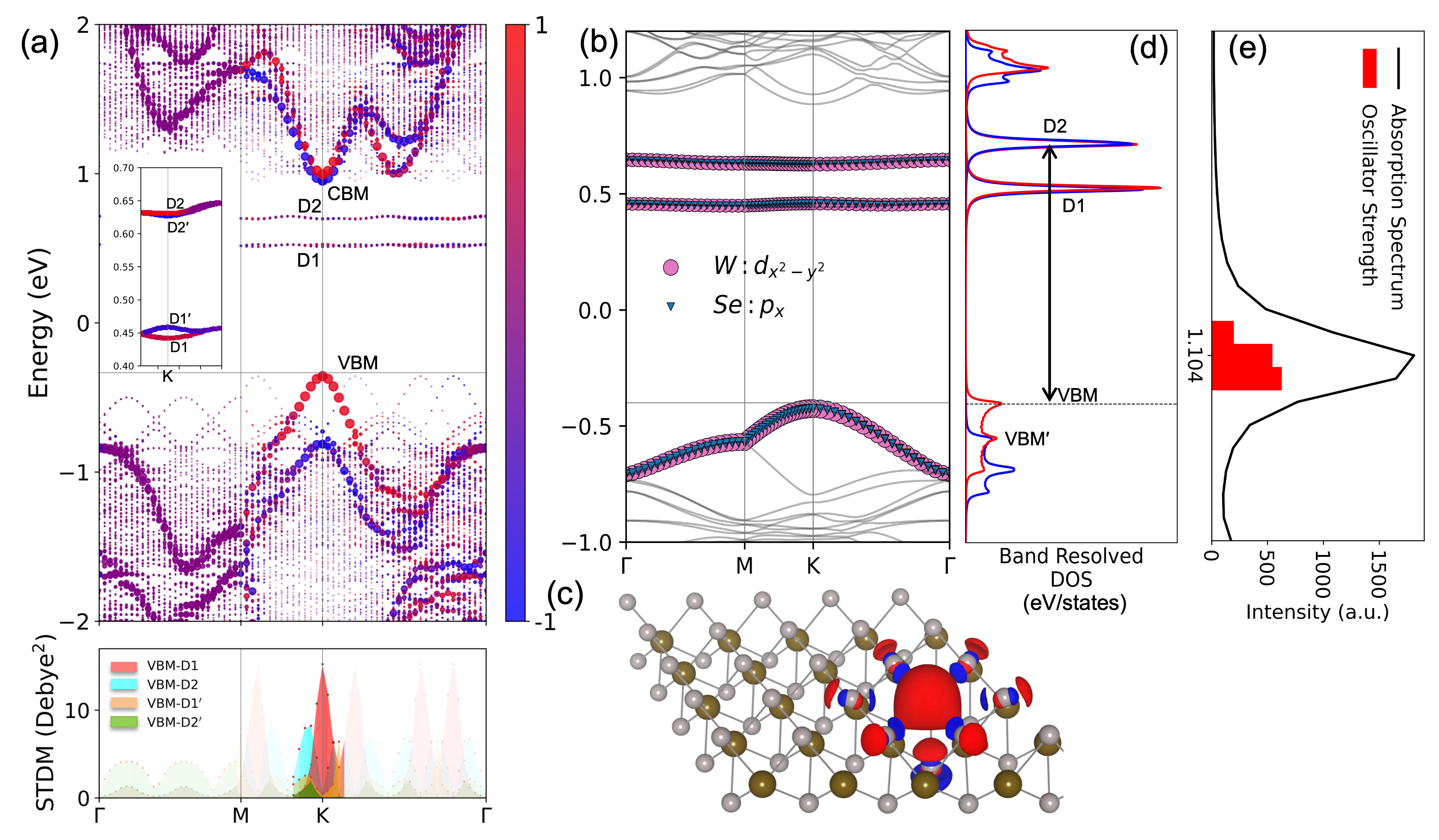}
 \caption{
 Electronic band structure with spin projection and defect-induced optical properties of monolayer WSe$_2$.
(a) Unfolded band structure of a $5\times5\times1$ WSe$_2$ monolayer supercell containing a single Se point defect, showing the spin-split conduction and valence bands near the K valley. Localized mid-gap defect states, D1 and D2, are likewise spin-split around the K valley. The lower panel shows the calculated spin-allowed squared transition dipole moments (STDMs) between the valence band (VB) and the defect states (D1 and D2) along the high-symmetry path of the Brillouin zone.
(b) Orbital-projected band structure of the VB and defect levels, showing comparable contributions from W-d and Se-p orbitals to the defect states.
(c) Real-space charge density distribution associated with the Se defect.
(d) Band-resolved partial density of states (PDOS) for the VB, D1, and D2 bands, colored according to their spin projections. Black Arrow indicates the spin-allowed optical emission from D2 $\rightarrow$ VB.
(e) Simulated absorption spectrum and oscillator strengths, showing dominant D2 defect-induced optical activity near 1.1 eV.}
\label{theory}
\end{figure*}

The microscopic origin of the defect emission was investigated using spin-polarized density functional theory (DFT) calculations on monolayer WSe$_2$ containing one representative Se vacancy (Se$_{\mathrm{vac}}$). Figure~\ref{theory} summarizes the electronic structure with the corresponding spin-resolved optical transitions for one of the Se$_{\mathrm{vac}}$ configuration. In pristine WSe$_2$, the valence and conduction bands exhibit spin splitting of 33 meV and 170 meV, respectively in our calculations due to strong spin-orbit coupling (SOC) near the K valley. However, the introduction of Se vacancy generates a set of localized electronic states within the band gap. Two defect manifolds, labeled D1 and D2, are consistently obtained at different Se defect concentrations designed on a WSe$_2$ supercell (see ESI). This indicates that these in-gap states are an intrinsic consequence of the Se vacancy and the associated breaking of W-Se bonds, rather than a structural artifact.

To elucidate the optical activity of the defect states, the lower panel of Fig. \ref{theory}(a) presents the calculated squared transition dipole moments (STDM) along the high-symmetry path as, $\boldsymbol{\mu}_{k}^2
=
\frac{i\hbar}{(\epsilon_{f,k}-\epsilon_{i,k})m}
|\left\langle
\psi_{f,k}
\middle|
\mathbf{p}
\middle|
\psi_{i,k}
\right\rangle|^2$, 
where $\epsilon_{i,k}$ and $\epsilon_{f,k}$ denote the initial and final state energies, $\psi_{i,k}$ and $\psi_{f,k}$ are the corresponding Bloch wavefunctions, and $\mathbf{p}$ is the momentum operator for direct vertical transitions. As shown in Fig.~\ref{theory}(b), both the valence band (VB) edge and the defect states predominantly originate from hybridized W-$d_{x^2-y^2}$ and Se-$p_x$ orbitals. The similarity in their orbital character results in a large transition dipole matrix element, enabling strong dipole-allowed optical transitions between defect states and the valence band maximum (VBM). Although, defect energy levels are nearly degenerate, SOC and the local symmetry breaking lift this degeneracy, giving rise to spin-split defect bands across certain parts of the Brillouin zone. For the considered case of 2\% defect, the spin split defect bands around $K$ are shown in the inset of Fig.~\ref{theory}(a). Therefore, the enhanced STDM around $K$ valley in the lower panel of Fig.~\ref{theory}(a) demonstrates strong momentum coupling between the VBM state and the corresponding D1 and D2 states of same spin color, indicating their active participation in spin-selective optical transitions. In contrast, the VBM$\rightarrow$D1$^\prime$ and
VBM$\rightarrow$D2$^\prime$ transitions, which involve states with
opposite spin character, exhibit weak coupling in the STDM calculations. Unlike the primary exciton nature of pristine WSe$_2$, the transitions involving the VB and defect labels are not concentrated in a single K-valley due to the localized nature of these defect states, which is evident from both their weak band dispersion (Fig.~\ref{theory}(a)) and the real-space charge density distributions (Fig.~\ref{theory} (c)). As shown in Fig.~\ref{theory}(c), the wavefunctions are strongly confined around the vacancy and its neighboring W atoms, whereas they span across the momentum-space, allowing optical transitions to involve multiple $k$-points across the BZ. As shown in ESI , the defect bands become more dispersive with a smaller WSe$_2$ supercell (\textit{i.e.} higher defect concentrations), where we have higher probability of neighboring defect interactions.

Variations in nearby impurities, dielectric screening, and
defect--defect interactions can modify the energies of localized defect
states. In an experimental sample, the measured defect-related PL band
is therefore likely to represent an inhomogeneous ensemble of
energetically closer, but non-identical, localized centres rather than
a single transition between two sharply defined electronic levels.

The spin-resolved density of states shown in Fig.~\ref{theory}(d)
reveals several energetically closely spaced transitions between the
valence-band states and the vacancy-induced in-gap states. The energy
separations between peaks having the same spin projection provide the
corresponding transition energies. The calculated oscillator strengths,
obtained from the Casida equation for the excitation-coupling matrix as
implemented in our \textit{ab initio} code\cite{sander2017macroscopic},
are shown in Fig.~\ref{theory}(e). These calculations demonstrate that
the vacancy-induced states can participate in optically active
transitions with the valence band. The reverse processes correspond to
the defect-related radiative recombination observed experimentally.

Because the defect states are spatially localized, their wavefunctions
are not associated with a sharply defined crystal momentum.
Localization therefore relaxes the strict momentum-selection rules
applicable to extended Bloch states and permits optical transitions
involving a broader range of momentum components. In addition, several
closely spaced defect-related transitions may overlap within the
experimental spectral resolution which
can produce a broad, inhomogeneously broadened defect-emission band
rather than a few spectrally resolved discrete transitions.

The variations in the local environment can also modify the
transition dipole moments, carrier-capture efficiencies, and radiative
and non-radiative recombination rates of individual defect centres\cite{ayari2018radiative,li2019defect}.
Consequently, the ensemble emission may exhibit a broad distribution of
recombination times, consistent with the experimentally observed
non-exponential decay. The comparatively large oscillator strengths calculated for
spin-allowed transitions indicate that these channels can contribute
strongly to the prompt radiative recombination and may therefore be
associated with the dominant short-lived exponential PL component. Defect-induced localization and symmetry breaking can relax the strict
momentum-selection rules of the pristine crystal, whereas spin--orbit
coupling can introduce mixing between states of different spin
character. These effects make both momentum-forbidden and
spin-forbidden defect-assisted transitions weakly allowed. The resulting
pathways are expected to have lower transition probabilities than the
direct spin-allowed channels and may therefore contribute to the slower,
long-lived component of the PL decay
\cite{yan2025defect,ahmed2023bright}.

The calculations therefore support the presence of multiple
defect-assisted optical pathways with a broad range of transition
probabilities. In combination with the variation in the local
environment of individual defects, these pathways provide a plausible
microscopic basis for long distributed recombination rates. Accordingly, the
long-lived power-law PL component is consistent with an inhomogeneous
ensemble of weakly allowed defect-assisted transitions, whereas the
short-lived exponential contribution is likely dominated by more
strongly allowed radiative channels.

\section{Methods}
\subsection{Sample preparation}
WSe$_2$ sample was exfoliated from commercially bought (from 2D Semiconductors) bulk sample using scotch-tape method and tranferred to on {90 nm} SiO$_2$/Si substrates. Monolayer was first identified by inspecting colour contrast of the flake under a otpical microscope.
\subsection{PL and Raman Measurements}
PL spectrum measurements at low temperature (below 70 K) were done using a home-built confocal microscope. The 525 nm laser was beam expanded and passed through objective (NA-0.82,working distance = 0.64 mm,focal length=2.88 mm) and focussed onto monolayer flake.
The spotsize of laser is as 447 nm spotsize from spotsize equation \cite{meinhart2003theory}. The sample and objective is in cryostat chamber (attoDRY800 closed Helium cryostat with base temperature of 4 kelvin on sample). The collected PL was filtered using a dichroic mirror and 550 long pass filter before focussing to Spectrometer(SpectraPro$*$- SP2150 spectrograph from  Teledyne Princeton Instruments)  The grating of 300 lines per mm was chosen for spectral measurements. The sample was located using white light imaging setup using a lens of 150 mm focal length and a webcam (Logitech Webcam).

Raman measurement was taken using confocal Raman spectrometer (Renishaw, inVia Reflex). PL spectrum (above 77K) was also taken using same setup using liquid nitrogen-cooled Linkam cryostat. 

\subsection{ Lifetime Measurements}
The same homebuilt setup was used for lifetime measurements. For time resolved measurements, laser was operated in pulsed mode of operation. The digital trigger was given to laser for pulsed operation using a pulse generator (Swabian Pulse Streamer with 5 ns on time out of  50  $\mu\text{s}$ pulse period). PL was redirected onto 50 $\mu\text{m}$ multimode fibre connected to SPAD (Excilitas-SPCM-AQRH-14-FC) using a flipmirror. The TTL pulses from SPAD is fed to a Time-Correlation unit (Swabian TimeTager 20). The timetagger was operated in histogram mode, where a histogram of the total delay time between the laser pulse and the emitted photons is created along the $x$-axis, covering $50~\mu\mathrm{s}$ with a resolution of $1~\mathrm{ns}$.
 The number of counts corresponding to each delay time is recorded in y axis. The histogram is averaged for each laser pulse. The total measurement time is $300~\mathrm{s}$, which corresponds to $\frac{300~\mathrm{s}}{50~\mu\mathrm{s}} = 6 \times 10^{6}$ averages.

\subsection{Computational Methods}
First-principles calculations were performed within the framework of density functional theory (DFT) using the projector augmented-wave (PAW) method as implemented in the VASP Package. The exchange-correlation interaction was treated using the generalized gradient approximation (GGA) within the PBE functional. Spin-orbit coupling (SOC) was explicitly included in all electronic structure calculations due to the relativistic effects associated with W-based TMDCs. A plane-wave energy cutoff of 420 eV was employed to ensure convergence of total energies and electronic properties. Structural relaxations were carried out until the Hellmann-Feynman forces on each atom were less than 10$^{-2}$ eV/{\AA}, while the total energy convergence criterion was set to 10$^{-6}$ eV.\\
Monolayer WSe$_2$ was modeled using a periodically repeated supercell geometry with a vacuum spacing of approximately 20 {\AA} along the out-of-plane direction to eliminate spurious interlayer interactions. A systematic study with consecutive $4\times4\times1$ and, $5\times5\times1$ supercells containing one Se vacancy, henceforth denoted as Se$_{vac}$ was constructed to minimize defect-defect interactions arising from periodic boundary conditions. The Brillouin zone integration was performed using a $\Gamma$-centered k-mesh of $2\times2\times1$  appropriate for the supercell size, while denser k-point sampling, $5\times5\times1$  was used for optical property calculations. Atomic positions and lattice parameters were fully optimized before subsequent electronic and optical analysis.

\section{Conclusion}

In summary, we investigated the recombination dynamics of localized
defect emission in monolayer WSe$_2$ using temperature- and
excitation-fluence-dependent time-resolved photoluminescence. The
defect-related emission contains a dominant sub-nanosecond exponential
component together with two weak, long-lived power-law channels
characterized by timescales ranging from a few nanoseconds to several
hundred nanoseconds. The power-law relaxation can be represented by
continuous distributions of recombination lifetimes, consistent with an
inhomogeneous ensemble of localized states experiencing different local
dielectric and electrostatic environment.

The temperature-dependent rates associated with both long-lived
channels are described by a Bose--Einstein phonon-occupation model,
supporting phonon-assisted detrapping. The extracted characteristic
energy scales are close to 60~meV and are consistent with a possible
two-optical-phonon-assisted relaxation process, although the specific
phonon pathway cannot be uniquely identified from the fit alone.
Spin-resolved electronic-structure and optical-transition calculations
for the thermodynamically favourable native Se-vacancy configuration
further reveal localized, optically active in-gap states and additional
weakly allowed spin and momentum forbidden  recombination pathways.

\section*{Author Contributions}

I.T. conceptualised the project, designed and performed all experiments,
analysed and intepreted the experimental data, and wrote the original manuscript.
S.J. performed the density functional theory calculations, analysed the
computational results, and wrote the corresponding section in the
manuscript. M.D. prepared the monolayer WSe$_2$ flakes by mechanical
exfoliation. B.R.K.N. supervised the computational component of the
project and contributed to scientific discussions    that supported the interpretation of the
project .V.P.B. supervised the experimental component, edited the manuscript and  contributed to the
interpretation and discussion of the broader implications of the
results.
\section*{Conflict of Interests}
The authors declare no competing financial or non-financial
interests.
\section*{Data Availability}
The data for this article will be uploaded to a standard public
repository upon acceptance, in accordance with the journal
guidelines.

\section*{Acknowledgment}
V. P. B. acknowledges the
financial support from DST QUEST grant DST/ICPS/QuST/
Theme-2/Q35. I.T acknowledges the
financial support from DST INSPIRE FELLOWSHIP.

\balance

\bibliography{rsc} 

@article{ayari2018radiative,
  title={Radiative lifetime of localized excitons in transition-metal dichalcogenides},
  author={Ayari, Sabrine and Smiri, Adlen and Hichri, Aida and Jaziri, Sihem and Amand, Thierry},
  journal={Physical Review B},
  volume={98},
  number={20},
  pages={205430},
  year={2018},
  publisher={APS}
}

@article{li2018revealing,
  title={Revealing the biexciton and trion-exciton complexes in BN encapsulated WSe2},
  author={Li, Zhipeng and Wang, Tianmeng and Lu, Zhengguang and Jin, Chenhao and Chen, Yanwen and Meng, Yuze and Lian, Zhen and Taniguchi, Takashi and Watanabe, Kenji and Zhang, Shengbai and others},
  journal={Nature communications},
  volume={9},
  number={1},
  pages={3719},
  year={2018},
  publisher={Nature Publishing Group UK London}
}

@article{liu2019gate,
  title={Gate tunable dark trions in monolayer WSe 2},
  author={Liu, Erfu and van Baren, Jeremiah and Lu, Zhengguang and Altaiary, Mashael M and Taniguchi, Takashi and Watanabe, Kenji and Smirnov, Dmitry and Lui, Chun Hung},
  journal={Physical review letters},
  volume={123},
  number={2},
  pages={027401},
  year={2019},
  publisher={APS}
}

@article{godde2016exciton,
  title={Exciton and trion dynamics in atomically thin MoSe 2 and WSe 2: Effect of localization},
  author={Godde, T and Schmidt, D and Schmutzler, J and A{\ss}mann, M and Debus, J and Withers, F and Alexeev, EM and Del Pozo-Zamudio, O and Skrypka, OV and Novoselov, KS and others},
  journal={Physical Review B},
  volume={94},
  number={16},
  pages={165301},
  year={2016},
  publisher={APS}
}

@article{wang2014valley,
  title={Valley dynamics probed through charged and neutral exciton emission in monolayer WSe 2},
  author={Wang, Gamg and Bouet, Louis and Lagarde, Delphine and Vidal, Ma{\"e}l and Balocchi, Andrea and Amand, Thierry and Marie, Xavier and Urbaszek, Bernhard},
  journal={Physical Review B},
  volume={90},
  number={7},
  pages={075413},
  year={2014},
  publisher={APS}
}

@article{meinhart2003theory,
  title={The theory of diffraction-limited resolution in microparticle image velocimetry},
  author={Meinhart, Carl D and Wereley, Steven T},
  journal={Measurement science and technology},
  volume={14},
  number={7},
  pages={1047},
  year={2003},
  publisher={IOP Publishing}
}

@article{sahin2013anomalous,
  title={Anomalous Raman spectra and thickness-dependent electronic properties of WSe 2},
  author={Sahin, H and Tongay, Sefaattin and Horzum, S and Fan, W and Zhou, J and Li, J and Wu, J and Peeters, FM},
  journal={Physical Review B—Condensed Matter and Materials Physics},
  volume={87},
  number={16},
  pages={165409},
  year={2013},
  publisher={APS}
}

@article{le2015spin,
  title={Spin--orbit coupling in the band structure of monolayer WSe2},
  author={Le, Duy and Barinov, Alexei and Preciado, Edwin and Isarraraz, Miguel and Tanabe, Iori and Komesu, Takashi and Troha, Conrad and Bartels, Ludwig and Rahman, Talat S and Dowben, Peter A},
  journal={Journal of Physics: Condensed Matter},
  volume={27},
  number={18},
  pages={182201},
  year={2015},
  publisher={IOP Publishing}
}

@article{kosmider2013large,
  title={Large spin splitting in the conduction band of transition metal dichalcogenide monolayers},
  author={Ko{\'s}mider, K and Gonz{\'a}lez, Jhon W and Fern{\'a}ndez-Rossier, Joaqu{\i}n},
  journal={Physical Review B},
  volume={88},
  number={24},
  pages={245436},
  year={2013},
  publisher={APS}
}

@article{ren2023measurement,
  title={Measurement of the conduction band spin-orbit splitting in WSe 2 and WS 2 monolayers},
  author={Ren, Lei and Robert, Cedric and Dery, Hanan and He, Minhao and Li, Pengke and Van Tuan, Dinh and Renucci, Pierre and Lagarde, Delphine and Taniguchi, Takashi and Watanabe, Kenji and others},
  journal={Physical Review B},
  volume={107},
  number={24},
  pages={245407},
  year={2023},
  publisher={APS}
}

@article{kapuscinski2021rydberg,
  title={Rydberg series of dark excitons and the conduction band spin-orbit splitting in monolayer WSe2},
  author={Kapu{\'s}ci{\'n}ski, Piotr and Delhomme, Alex and Vaclavkova, Diana and Slobodeniuk, Artur O and Grzeszczyk, Magdalena and Bartos, Miroslav and Watanabe, Kenji and Taniguchi, Takashi and Faugeras, Cl{\'e}ment and Potemski, Marek},
  journal={Communications Physics},
  volume={4},
  number={1},
  pages={186},
  year={2021},
  publisher={Nature Publishing Group UK London}
}

@article{ye2018efficient,
  title={Efficient generation of neutral and charged biexcitons in encapsulated WSe2 monolayers},
  author={Ye, Ziliang and Waldecker, Lutz and Ma, Eric Yue and Rhodes, Daniel and Antony, Abhinandan and Kim, Bumho and Zhang, Xiao-Xiao and Deng, Minda and Jiang, Yuxuan and Lu, Zhengguang and others},
  journal={Nature communications},
  volume={9},
  number={1},
  pages={3718},
  year={2018},
  publisher={Nature Publishing Group UK London}
}

@article{zhang2015experimental,
  title={Supplemental Material of Experimental evidence for dark excitons in monolayer WSe 2},
  author={Zhang, Xiao-Xiao and You, Yumeng and Zhao, Shu Yang Frank and Heinz, Tony F},
  journal={Physical review letters},
  volume={115},
  number={25},
  pages={257403},
  year={2015},
  publisher={APS}
}

@article{wagner2021trap,
  title={Trap induced long exciton intervalley scattering and population lifetime in monolayer WSe2},
  author={Wagner, Julian and Kuhn, Henning and Bernhardt, Robin and Zhu, Jingyi and Van Loosdrecht, Paul HM},
  journal={2D Materials},
  volume={8},
  number={3},
  pages={035018},
  year={2021},
  publisher={IOP Publishing}
}

@article{kipczak2024impact,
  title={Impact of temperature on the brightening of neutral and charged dark excitons in WSe2 monolayer},
  author={Kipczak, {\L}ucja and Zawadzka, Natalia and Jana, Dipankar and Antoniazzi, Igor and Grzeszczyk, Magdalena and Zinkiewicz, Ma{\l}gorzata and Watanabe, Kenji and Taniguchi, Takashi and Potemski, Marek and Faugeras, Cl{\'e}ment and others},
  journal={Nanophotonics},
  volume={13},
  number={26},
  pages={4743--4749},
  year={2024},
  publisher={De Gruyter}
}

@article{davila2024temperature,
  title={Temperature and power-dependent photoluminescence spectroscopy in suspended WSe2 monolayer},
  author={Davila, Yuset Guerra and Silva, Francisco WN and Oliveira, Maykol CD and Yu, Zhuohang and Carvalho, Thais CV and dos Santos, Clenilton C and Souza Filho, Antonio G and Terrones, Mauricio and Alencar, Rafael S and Viana, Bartolomeu C},
  journal={Journal of Physics D: Applied Physics},
  volume={57},
  number={16},
  pages={165304},
  year={2024},
  publisher={IOP Publishing}
}

@article{tongay2013defects,
  title={Defects activated photoluminescence in two-dimensional semiconductors: interplay between bound, charged and free excitons},
  author={Tongay, Sefaattin and Suh, Joonki and Ataca, Can and Fan, Wen and Luce, Alexander and Kang, Jeong Seuk and Liu, Jonathan and Ko, Changhyun and Raghunathanan, Rajamani and Zhou, Jian and others},
  journal={Scientific reports},
  volume={3},
  number={1},
  pages={2657},
  year={2013},
  publisher={Nature Publishing Group UK London}
}

@article{robert2017fine,
  title={Fine structure and lifetime of dark excitons in transition metal dichalcogenide monolayers},
  author={Robert, C{\'e}dric and Amand, Thierry and Cadiz, Fabian and Lagarde, Delphine and Courtade, Emmanuel and Manca, Marco and Taniguchi, Takashi and Watanabe, Kenji and Urbaszek, Bernhard and Marie, Xavier},
  journal={Physical review B},
  volume={96},
  number={15},
  pages={155423},
  year={2017},
  publisher={APS}
}

@article{zhou2017probing,
  title={Probing dark excitons in atomically thin semiconductors via near-field coupling to surface plasmon polaritons},
  author={Zhou, You and Scuri, Giovanni and Wild, Dominik S and High, Alexander A and Dibos, Alan and Jauregui, Luis A and Shu, Chi and De Greve, Kristiaan and Pistunova, Kateryna and Joe, Andrew Y and others},
  journal={Nature nanotechnology},
  volume={12},
  number={9},
  pages={856--860},
  year={2017},
  publisher={Nature Publishing Group UK London}
}

@article{wang2017plane,
  title={In-plane propagation of light in transition metal dichalcogenide monolayers: optical selection rules},
  author={Wang, Gang and Robert, C{\'e}dric and Glazov, Mikhail M and Cadiz, Fabian and Courtade, Emmanuel and Amand, Thierry and Lagarde, Delphine and Taniguchi, Takashi and Watanabe, Kenji and Urbaszek, Bernhard and others},
  journal={Physical review letters},
  volume={119},
  number={4},
  pages={047401},
  year={2017},
  publisher={APS}
}

@article{lo2022plasmonic,
  title={Plasmonic nanocavity induced coupling and boost of dark excitons in monolayer WSe2 at room temperature},
  author={Lo, Tsz Wing and Chen, Xiaolin and Zhang, Zhedong and Zhang, Qiang and Leung, Chi Wah and Zayats, Anatoly V and Lei, Dangyuan},
  journal={Nano Letters},
  volume={22},
  number={5},
  pages={1915--1921},
  year={2022},
  publisher={ACS Publications}
}

@article{kim1994thermodynamics,
  title={Thermodynamics of biexcitons in a GaAs quantum well},
  author={Kim, JC and Wake, DR and Wolfe, JP},
  journal={Physical Review B},
  volume={50},
  number={20},
  pages={15099},
  year={1994},
  publisher={APS}
}

@article{you2015observation,
  title={Observation of biexcitons in monolayer WSe 2},
  author={You, Yumeng and Zhang, Xiao-Xiao and Berkelbach, Timothy C and Hybertsen, Mark S and Reichman, David R and Heinz, Tony F},
  journal={Nature Physics},
  volume={11},
  number={6},
  pages={477--481},
  year={2015},
  publisher={Nature Publishing Group UK London}
}

@article{phillips1992biexciton,
  title={Biexciton creation and recombination in a GaAs quantum well},
  author={Phillips, RT and Lovering, DJ and Denton, GJ and Smith, GW},
  journal={Physical Review B},
  volume={45},
  number={8},
  pages={4308},
  year={1992},
  publisher={APS}
}

@article{huang2016probing,
  title={Probing the origin of excitonic states in monolayer WSe 2},
  author={Huang, Jiani and Hoang, Thang B and Mikkelsen, Maiken H},
  journal={Scientific reports},
  volume={6},
  number={1},
  pages={22414},
  year={2016},
  publisher={Nature Publishing Group UK London}
}

@article{barbone2018charge,
  title={Charge-tuneable biexciton complexes in monolayer WSe2},
  author={Barbone, Matteo and Montblanch, Alejandro R-P and Kara, Dhiren M and Palacios-Berraquero, Carmen and Cadore, Alisson R and De Fazio, Domenico and Pingault, Benjamin and Mostaani, Elaheh and Li, Han and Chen, Bin and others},
  journal={Nature communications},
  volume={9},
  number={1},
  pages={3721},
  year={2018},
  publisher={Nature Publishing Group UK London}
}

@article{tang2019long,
  title={Long valley lifetime of dark excitons in single-layer WSe2},
  author={Tang, Yanhao and Mak, Kin Fai and Shan, Jie},
  journal={Nature communications},
  volume={10},
  number={1},
  pages={4047},
  year={2019},
  publisher={Nature Publishing Group UK London}
}

@misc{zheng2019first,
      title={First Principles Study of Intrinsic and Extrinsic Point Defects in Monolayer WSe2}, 
      author={Yu Jie Zheng and Su Ying Quek},
      year={2019},
      eprint={1901.05238},
      archivePrefix={arXiv},
      primaryClass={cond-mat.mtrl-sci},
      url={https://arxiv.org/abs/1901.05238}, 
}

@article{tosun2016air,
  title={Air-stable n-doping of WSe2 by anion vacancy formation with mild plasma treatment},
  author={Tosun, Mahmut and Chan, Leslie and Amani, Matin and Roy, Tania and Ahn, Geun Ho and Taheri, Peyman and Carraro, Carlo and Ager, Joel W and Maboudian, Roya and Javey, Ali},
  journal={ACS nano},
  volume={10},
  number={7},
  pages={6853--6860},
  year={2016},
  publisher={ACS Publications}
}

@article{nguyen2021gate,
  title={Gate-Tunable Magnetism via Resonant Se-Vacancy Levels in WSe2},
  author={Nguyen, Tuan Dung and Jiang, Jinbao and Song, Bumsub and Tran, Minh Dao and Choi, Wooseon and Kim, Ji Hee and Kim, Young-Min and Duong, Dinh Loc and Lee, Young Hee},
  journal={Advanced Science},
  volume={8},
  number={24},
  pages={2102911},
  year={2021},
  publisher={Wiley Online Library}
}

@article{lin2015three,
  title={Three-fold rotational defects in two-dimensional transition metal dichalcogenides},
  author={Lin, Yung-Chang and Bj{\"o}rkman, Torbj{\"o}rn and Komsa, Hannu-Pekka and Teng, Po-Yuan and Yeh, Chao-Hui and Huang, Fei-Sheng and Lin, Kuan-Hung and Jadczak, Joanna and Huang, Ying-Sheng and Chiu, Po-Wen and others},
  journal={Nature communications},
  volume={6},
  number={1},
  pages={6736},
  year={2015},
  publisher={Nature Publishing Group UK London}
}

@article{parto2021defect,
  title={Defect and strain engineering of monolayer WSe2 enables site-controlled single-photon emission up to 150 K},
  author={Parto, Kamyar and Azzam, Shaimaa I and Banerjee, Kaustav and Moody, Galan},
  journal={Nature communications},
  volume={12},
  number={1},
  pages={3585},
  year={2021},
  publisher={Nature Publishing Group UK London}
}

@article{qian2020defect,
  title={Defect creation in WSe 2 with a microsecond photoluminescence lifetime by focused ion beam irradiation},
  author={Qian, Qingkai and Peng, Lintao and Perea-Lopez, Nestor and Fujisawa, Kazunori and Zhang, Kunyan and Zhang, Xiaotian and Choudhury, Tanushree H and Redwing, Joan M and Terrones, Mauricio and Ma, Xuedan and others},
  journal={Nanoscale},
  volume={12},
  number={3},
  pages={2047--2056},
  year={2020},
  publisher={Royal Society of Chemistry}
}

@article{moody2018microsecond,
  title={Microsecond valley lifetime of defect-bound excitons in monolayer WSe 2},
  author={Moody, Galan and Tran, Kha and Lu, Xiaobo and Autry, Travis and Fraser, James M and Mirin, Richard P and Yang, Li and Li, Xiaoqin and Silverman, Kevin L},
  journal={Physical review letters},
  volume={121},
  number={5},
  pages={057403},
  year={2018},
  publisher={APS}
}

@article{choi2023chemical,
  title={Is chemical vapor deposition of monolayer WSe2 comparable to other synthetic routes?},
  author={Choi, Soo Ho and Yang, Sang-Hyeok and Park, Sehwan and Cho, Byeong Wook and Nguyen, Tuan Dung and Kim, Jung Ho and Kim, Young-Min and Kim, Ki Kang and Lee, Young Hee},
  journal={APL Materials},
  volume={11},
  number={11},
  pages = {111124},
  year={2023},
  publisher={AIP Publishing}
}

@article{andersen2003temperature,
  title={Temperature concepts for small, isolated systems; 1/t decay and radiative cooling},
  author={Andersen, Jens Ulrik and Bonderup, E and Hansen, K and Hvelplund, P and Liu, B and Pedersen, UV and Tomita, S},
  journal={The European Physical Journal D-Atomic, Molecular, Optical and Plasma Physics},
  volume={24},
  number={1},
  pages={191--196},
  year={2003},
  publisher={Springer}
}

@article{Mouri_2014,
   title={Nonlinear photoluminescence in atomically thin layered<mml:math xmlns:mml=“http://www.w3.org/1998/Math/MathML”><mml:msub><mml:mi>WSe</mml:mi><mml:mn>2</mml:mn></mml:msub></mml:math>arising from diffusion-assisted exciton-exciton annihilation},
   volume={90},
   ISSN={1550-235X},
   url={http://dx.doi.org/10.1103/PhysRevB.90.155449},
   DOI={10.1103/physrevb.90.155449},
   number={15},
   journal={Physical Review B},
   pages={155449},
   publisher={American Physical Society (APS)},
   author={Mouri, Shinichiro and Miyauchi, Yuhei and Toh, Minglin and Zhao, Weijie and Eda, Goki and Matsuda, Kazunari},
   year={2014},
   month=Oct }

@article{PhysRevA.77.042719,
  title = {Long-time deviations from exponential decay for inverse-square potentials},
  author = {Martorell, J. and Muga, J. G. and Sprung, D. W. L.},
  journal = {Phys. Rev. A},
  volume = {77},
  issue = {4},
  pages = {042719},
  numpages = {9},
  year = {2008},
  month = {Apr},
  publisher = {American Physical Society},
  doi = {10.1103/PhysRevA.77.042719},
  url = {https://link.aps.org/doi/10.1103/PhysRevA.77.042719}
}

@article{houel2015autocorrelation,
  title={Autocorrelation analysis for the unbiased determination of power-law exponents in single-quantum-dot blinking},
  author={Houel, Julien and Doan, Quang T and Cajgfinger, Thomas and Ledoux, Gilles and Amans, David and Aubret, Antoine and Dominjon, Agnes and Ferriol, Sylvain and Barbier, R{\'e}mi and Nasilowski, Michel and others},
  journal={ACS nano},
  volume={9},
  number={1},
  pages={886--893},
  year={2015},
  publisher={ACS Publications}
}

@article{yuan2024shallow,
  title={Shallow defects and variable photoluminescence decay times up to 280 $\mu$s in triple-cation perovskites},
  author={Yuan, Ye and Yan, Genghua and Dreessen, Chris and Rudolph, Toby and H{\"u}lsbeck, Markus and Klingebiel, Benjamin and Ye, Jiajiu and Rau, Uwe and Kirchartz, Thomas},
  journal={Nature materials},
  volume={23},
  number={3},
  pages={391--397},
  year={2024},
  publisher={Nature Publishing Group UK London}
}

@article{dass2019ultra,
  title={Ultra-long lifetimes of single quantum emitters in monolayer WSe2/hBN heterostructures},
  author={Dass, Chandriker Kavir and Khan, Mahtab A and Clark, Genevieve and Simon, Jeffrey A and Gibson, Ricky and Mou, Shin and Xu, Xiaodong and Leuenberger, Michael N and Hendrickson, Joshua R},
  journal={Advanced Quantum Technologies},
  volume={2},
  number={5-6},
  pages={1900022},
  year={2019},
  publisher={Wiley Online Library}
}

@article{barja2019identifying,
  title={Identifying substitutional oxygen as a prolific point defect in monolayer transition metal dichalcogenides},
  author={Barja, Sara and Refaely-Abramson, Sivan and Schuler, Bruno and Qiu, Diana Y and Pulkin, Artem and Wickenburg, Sebastian and Ryu, Hyejin and Ugeda, Miguel M and Kastl, Christoph and Chen, Christopher and others},
  journal={Nature communications},
  volume={10},
  number={1},
  pages={3382},
  year={2019},
  publisher={Nature Publishing Group UK London}
}

@article{lin2018realizing,
  title={Realizing large-scale, electronic-grade two-dimensional semiconductors},
  author={Lin, Yu-Chuan and Jariwala, Bhakti and Bersch, Brian M and Xu, Ke and Nie, Yifan and Wang, Baoming and Eichfeld, Sarah M and Zhang, Xiaotian and Choudhury, Tanushree H and Pan, Yi and others},
  journal={ACS nano},
  volume={12},
  number={2},
  pages={965--975},
  year={2018},
  publisher={ACS Publications}
}

@article{zhang2017defect,
  title={Defect structure of localized excitons in a WSe 2 monolayer},
  author={Zhang, Shuai and Wang, Chen-Guang and Li, Ming-Yang and Huang, Di and Li, Lain-Jong and Ji, Wei and Wu, Shiwei},
  journal={Physical review letters},
  volume={119},
  number={4},
  pages={046101},
  year={2017},
  publisher={APS}
}

@article{akmaev2020nonexponential,
  author  = {Akmaev, M. A. and Kochiev, M. V. and Duleba, A. I.
             and Pugachev, M. V. and Kuntsevich, A. Yu.
             and Belykh, V. V.},
  title   = {Nonexponential Photoluminescence Dynamics in an
             Inhomogeneous Ensemble of Excitons in {WSe$_2$}
             Monolayers},
  journal = {JETP Letters},
  volume  = {112},
  pages   = {607--614},
  year    = {2020},
  doi     = {10.1134/S0021364020220063}
}

@article{zheng2019point,
  title={Point defects and localized excitons in 2D WSe2},
  author={Zheng, Yu Jie and Chen, Yifeng and Huang, Yu Li and Gogoi, Pranjal Kumar and Li, Ming-Yang and Li, Lain-Jong and Trevisanutto, Paolo E and Wang, Qixing and Pennycook, Stephen J and Wee, Andrew TS and others},
  journal={ACS nano},
  volume={13},
  number={5},
  pages={6050--6059},
  year={2019},
  publisher={ACS Publications}
}

@article{carin2024role,
  title={The Role of Chalcogen Vacancies in Single Photon Emission from Monolayer Tungsten Dichalcogenides},
  author={Carin Gavin, S and Zeman IV, Charles J and Dasgupta, Anushka and Liu, Yiying and Wu, Wenjing and Huang, Shengxi and Marks, Tobin J and Hersam, Mark C and Schatz, George C and Stern, Nathaniel P},
  journal={arXiv e-prints},
  pages={arXiv--2412},
  year={2024}
}

@article{kuno2000nonexponential,
  title={Nonexponential “blinking” kinetics of single CdSe quantum dots: A universal power law behavior},
  author={Kuno, Masaru and Fromm, David P and Hamann, Hendrik F and Gallagher, Alan and Nesbitt, David J},
  journal={The journal of chemical physics},
  volume={112},
  number={7},
  pages={3117--3120},
  year={2000},
  publisher={American Institute of Physics}
}

@article{erkensten2021dark,
  title={Dark exciton-exciton annihilation in monolayer WSe 2},
  author={Erkensten, Daniel and Brem, Samuel and Wagner, Koloman and Gillen, Roland and Perea-Caus{\'\i}n, Ra{\"u}l and Ziegler, Jonas D and Taniguchi, Takashi and Watanabe, Kenji and Maultzsch, Janina and Chernikov, Alexey and others},
  journal={Physical Review B},
  volume={104},
  number={24},
  pages={L241406},
  year={2021},
  publisher={APS}
}

@article{xu2024control,
author = {Xu, Haowen and Wang, Jiangcai and Liu, Huan and Chen, Shihong and Sun, Zejun and Wang, Chong and Han, Rui and Wang, Yong and Wang, Yutang and Wang, Zihao and Huang, Shuchun and Ma, Lingwei and Liu, Dameng},
title = {Control of Hybrid Exciton Lifetime in MoSe2/WS2 Moiré Heterostructures},
journal = {Advanced Science},
volume = {11},
number = {34},
pages = {2403127},
doi = {https://doi.org/10.1002/advs.202403127},
url = {https://advanced.onlinelibrary.wiley.com/doi/abs/10.1002/advs.202403127},
eprint = {https://advanced.onlinelibrary.wiley.com/doi/pdf/10.1002/advs.202403127},
year = {2024}
}

@article{martin2008observation,
  title={Observation of electron--hole puddles in graphene using a scanning single-electron transistor},
  author={Martin, Jens and Akerman, Nitzan and Ulbricht, G and Lohmann, T and Smet, JH v and Von Klitzing, K and Yacoby, Amir},
  journal={Nature physics},
  volume={4},
  number={2},
  pages={144--148},
  year={2008},
  publisher={Nature Publishing Group UK London}
}

@article{zhang2010origin,
  title={Origin of spatial charge inhomogeneity in graphene},
  author={Zhang, Yuanbo and Brar, Victor W and Girit, Caglar and Zettl, Alex and Crommie, Michael F},
  journal={Nature Physics},
  volume={6},
  number={1},
  pages={74},
  year={2010},
  publisher={Nature Publishing Group}
}

@article{rhodes2019disorder,
  title={Disorder in van der Waals heterostructures of 2D materials},
  author={Rhodes, Daniel and Chae, Sang Hoon and Ribeiro-Palau, Rebeca and Hone, James},
  journal={Nature materials},
  volume={18},
  number={6},
  pages={541--549},
  year={2019},
  publisher={Nature Publishing Group UK London}
}

@article{lopion2020temperature,
  title={Temperature dependence of photoluminescence lifetime of atomically-thin WSe2 layer},
  author={{\L}opion, Aleksandra and Goryca, Mateusz and Smole{\'n}ski, Tomasz and Oreszczuk, Kacper and Nogajewski, Karol and Molas, Maciej R and Potemski, Marek and Kossacki, Piotr},
  journal={Nanotechnology},
  volume={31},
  number={13},
  pages={135002},
  year={2020},
  publisher={IOP Publishing}
}

@article{hernandez2022strain,
  title={Strain control of hybridization between dark and localized excitons in a 2D semiconductor},
  author={Hern{\'a}ndez L{\'o}pez, Pablo and Heeg, Sebastian and Schattauer, Christoph and Kovalchuk, Sviatoslav and Kumar, Abhijeet and Bock, Douglas J and Kirchhof, Jan N and H{\"o}fer, Bianca and Greben, Kyrylo and Yagodkin, Denis and others},
  journal={Nature communications},
  volume={13},
  number={1},
  pages={7691},
  year={2022},
  publisher={Nature Publishing Group UK London}
}

@article{chaudhary2020origin,
  title={Origin of selective enhancement of sharp defect emission lines in monolayer WSe2 on rough metal substrate},
  author={Chaudhary, Raghav and Raghunathan, Varun and Majumdar, Kausik},
  journal={Journal of Applied Physics},
  volume={127},
  number={7},
  pages = {073105},
  year={2020},
  publisher={AIP Publishing}
}

@article{tonndorf2013photoluminescence,
  title   = {Photoluminescence emission and Raman response of monolayer
             MoS2, MoSe2, and WSe2},
  author  = {Tonndorf, P. and others},
  journal = {Optics Express},
  volume  = {21},
  pages   = {4908--4916},
  year    = {2013},
  doi     = {10.1364/OE.21.004908}
}

@article{wang2015ultrafast,
  title={Ultrafast dynamics of defect-assisted electron--hole recombination in monolayer MoS2},
  author={Wang, Haining and Zhang, Changjian and Rana, Farhan},
  journal={Nano letters},
  volume={15},
  number={1},
  pages={339--345},
  year={2015},
  publisher={ACS Publications}
}

@article{deng2020long,
  title={Long-range exciton transport and slow annihilation in two-dimensional hybrid perovskites},
  author={Deng, Shibin and Shi, Enzheng and Yuan, Long and Jin, Linrui and Dou, Letian and Huang, Libai},
  journal={Nature communications},
  volume={11},
  number={1},
  pages={664},
  year={2020},
  publisher={Nature Publishing Group UK London}
}

@article{liu2019neutral,
  title={Neutral and defect-induced exciton annihilation in defective monolayer WS 2},
  author={Liu, Huan and Wang, Chong and Liu, Dameng and Luo, Jianbin},
  journal={Nanoscale},
  volume={11},
  number={16},
  pages={7913--7920},
  year={2019},
  publisher={Royal Society of Chemistry}
}

@article{steinleitner2017direct,
  author  = {Steinleitner, Philipp and Merkl, Philipp and Nagler, Philipp
             and Mornhinweg, Joshua and Sch{\"u}ller, Christian and Korn,
             Tobias and Chernikov, Alexey and Huber, Rupert},
  title   = {Direct Observation of Ultrafast Exciton Formation in a
             Monolayer of WSe2},
  journal = {Nano Letters},
  volume  = {17},
  number  = {3},
  pages   = {1455--1460},
  year    = {2017},
  doi     = {10.1021/acs.nanolett.6b04422}
}

@article{li2018auger,
  title={The Auger process in multilayer WSe 2 crystals},
  author={Li, Yuanzheng and Shi, Jia and Chen, Heyu and Wang, Rui and Mi, Yang and Zhang, Cen and Du, Wenna and Zhang, Shuai and Liu, Zheng and Zhang, Qing and others},
  journal={Nanoscale},
  volume={10},
  number={37},
  pages={17585--17592},
  year={2018},
  publisher={Royal Society of Chemistry}
}

@article{shin2020ultrafast,
  title={Ultrafast Auger process in few-layer PtSe 2},
  author={Shin, Hee Jun and Bae, Seongkwang and Sim, Sangwan},
  journal={Nanoscale},
  volume={12},
  number={43},
  pages={22185--22191},
  year={2020},
  publisher={Royal Society of Chemistry}
}

@article{yan2025defect,
  title={Defect-Induced Spin Splitting Extends Charge Carrier Lifetimes in Anatase TiO2},
  author={Yan, Xiaodan and Han, Xiao and He, Jinlu},
  journal={The Journal of Physical Chemistry Letters},
  volume={16},
  number={18},
  pages={4428--4435},
  year={2025},
  publisher={ACS Publications}
}

@article{ahmed2023bright,
  title={Bright and efficient light-emitting devices based on 2D transition metal dichalcogenides},
  author={Ahmed, Tanveer and Zha, Jiajia and Lin, Kris KH and Kuo, Hao-Chung and Tan, Chaoliang and Lien, Der-Hsien},
  journal={Advanced Materials},
  volume={35},
  number={31},
  pages={2208054},
  year={2023},
  publisher={Wiley Online Library}
}

@article{sander2017macroscopic,
  title={Macroscopic dielectric function within time-dependent density functional theory—Real time evolution versus the Casida approach},
  author={Sander, Tobias and Kresse, Georg},
  journal={The Journal of Chemical Physics},
  volume={146},
  number={6},
  year={2017},
  pages = {064110},
  publisher={AIP Publishing}
}

@article{Zhao2013,
    author = {Zhao, Y. and Luo, X. and Li, H. and Zhang, J. and Araujo, P. T. and Gan, C. K. and Wu, J. and Zhang, H. and Quek, S. Y. and Dresselhaus, M. S. and Xiong, Q.},
    title = {Origin of the {Raman} {Breather} {Mode} in Layered {MoS}$_2$ and {WSe}$_2$},
    journal = {Phys. Rev. B},
    volume = {88},
    issue = {7},
    pages = {075317},
    numpages = {7},
    year = {2013},
    month = {Aug},
    publisher = {American Physical Society},
    doi = {10.1103/PhysRevB.88.075317},
    url = {https://link.aps.org/doi/10.1103/PhysRevB.88.075317}
}

@article{ziarkash2018comparative,
  title={Comparative study of afterpulsing behavior and models in single photon counting avalanche photo diode detectors},
  author={Ziarkash, Abdul Waris and Joshi, Siddarth Koduru and Stip{\v{c}}evi{\'c}, Mario and Ursin, Rupert},
  journal={Scientific reports},
  volume={8},
  number={1},
  pages={1--8},
  year={2018},
  publisher={Nature Publishing Group}
}

@article{palummo2015exciton,
  title={Exciton radiative lifetimes in two-dimensional transition metal dichalcogenides},
  author={Palummo, Maurizia and Bernardi, Marco and Grossman, Jeffrey C},
  journal={Nano letters},
  volume={15},
  number={5},
  pages={2794--2800},
  year={2015},
  publisher={ACS Publications}
}

@article{lu2015atomic,
  title={Atomic healing of defects in transition metal dichalcogenides},
  author={Lu, Junpeng and Carvalho, Alexandra and Chan, Xinhui Kim and Liu, Hongwei and Liu, Bo and Tok, Eng Soon and Loh, Kian Ping and Castro Neto, AH and Sow, Chorng Haur},
  journal={Nano letters},
  volume={15},
  number={5},
  pages={3524--3532},
  year={2015},
  publisher={ACS Publications}
}

@article{guerra2024temperature,
  title={Temperature and power-dependent photoluminescence spectroscopy in suspended WSe2 monolayer},
  author={Guerra Davila, Yuset and Silva, Francisco WN and Oliveira, Maykol CD and Yu, Zhuohang and Carvalho, Thais CV and dos Santos, Clenilton C and Souza Filho, Antonio G and Terrones, Mauricio and Alencar, Rafael S and Viana, Bartolomeu C},
  journal={Journal of Physics D: Applied Physics},
  volume={57},
  number={16},
  pages={165304},
  year={2024},
  publisher={IOP Publishing}
}

@book{widder1941laplace,
  author    = {David Vernon Widder},
  title     = {The Laplace Transform},
  publisher = {Princeton University Press},
  year      = {1941}
}

@article{inokuti1965influence,
  author  = {Inokuti, Mitio and Hirayama, Fumio},
  title   = {Influence of Energy Transfer by the Exchange Mechanism
             on Donor Luminescence},
  journal = {The Journal of Chemical Physics},
  volume  = {43},
  number  = {6},
  pages   = {1978--1989},
  year    = {1965},
  doi     = {10.1063/1.1697063}
}

@article{takagishi2007stretched,
  author  = {Takagishi, Shigenori and Hashimoto, J. and Nakayama, Masaaki},
  title   = {Stretched Exponential Profiles of Photoluminescence
             Decays Related to Localized States in
             InGaAsN/GaAs Single-Quantum Wells},
  journal = {Journal of Luminescence},
  volume  = {122--123},
  pages   = {753--755},
  year    = {2007},
  doi     = {10.1016/j.jlumin.2006.01.279}
}

@article{reshchikov2018two,
  author  = {Reshchikov, Michael A. and Makarov, Dmitry and
             Helava, Heikki and Morko{\c{c}}, Hadis},
  title   = {Two Yellow Luminescence Bands in Undoped GaN},
  journal = {Scientific Reports},
  volume  = {8},
  pages   = {8091},
  year    = {2018},
  doi     = {10.1038/s41598-018-26354-z}
}

@article{scher1975anomalous,
  author  = {Scher, Harvey and Montroll, Elliott W.},
  title   = {Anomalous Transit-Time Dispersion in Amorphous Solids},
  journal = {Physical Review B},
  volume  = {12},
  number  = {6},
  pages   = {2455--2477},
  year    = {1975},
  doi     = {10.1103/PhysRevB.12.2455}
}

@article{martin2008onedimensional,
  author  = {Martin, James E. and Shea-Rohwer, Lauren E.},
  title   = {A 1-D Model of the Photoluminescent Decay of ZnS
             Phosphors as a Function of Excitation Conditions},
  journal = {Journal of Luminescence},
  volume  = {128},
  number  = {9},
  pages   = {1407--1420},
  year    = {2008},
  doi     = {10.1016/j.jlumin.2008.01.019}
}

@article{LiQuantum2022,
author = {Li, Xinxin and Wang, Wei and Ma, Xuedan},
title = {Supplementary Information of Quantum Photon Sources in WSe2 Monolayers Induced by Weakly Localized Strain Fields},
journal = {The Journal of Physical Chemistry C},
volume = {126},
number = {47},
pages = {20057-20064},
year = {2022},
doi = {10.1021/acs.jpcc.2c06148},

URL = { 
    
        https://doi.org/10.1021/acs.jpcc.2c06148
    
    

},
eprint = { 
    
        https://doi.org/10.1021/acs.jpcc.2c06148
    
    

}

}

@article{Mainfluenceof2015,
  title = {Influences of Exciton Diffusion and Exciton-Exciton Annihilation on Photon Emission Statistics of Carbon Nanotubes},
  author = {Ma, Xuedan and Roslyak, Oleskiy and Duque, Juan G. and Pang, Xiaoying and Doorn, Stephen K. and Piryatinski, Andrei and Dunlap, David H. and Htoon, Han},
  journal = {Phys. Rev. Lett.},
  volume = {115},
  issue = {1},
  pages = {017401},
  numpages = {6},
  year = {2015},
  month = {Jul},
  publisher = {American Physical Society},
  doi = {10.1103/PhysRevLett.115.017401},
  url = {https://link.aps.org/doi/10.1103/PhysRevLett.115.017401}
}

@article{Pengbright2020,
  title = {Bright trion emission from semiconductor nanoplatelets},
  author = {Peng, Lintao and Otten, Matthew and Hazarika, Abhijit and Coropceanu, Igor and Cygorek, Moritz and Wiederrecht, Gary P. and Hawrylak, Pawel and Talapin, Dmitri V. and Ma, Xuedan},
  journal = {Phys. Rev. Mater.},
  volume = {4},
  issue = {5},
  pages = {056006},
  numpages = {7},
  year = {2020},
  month = {May},
  publisher = {American Physical Society},
  doi = {10.1103/PhysRevMaterials.4.056006},
  url = {https://link.aps.org/doi/10.1103/PhysRevMaterials.4.056006}
}

@article{
klimovquantization2020,
author = {V. I. Klimov  and A. A. Mikhailovsky  and D. W. McBranch  and C. A. Leatherdale  and M. G. Bawendi },
title = {Quantization of Multiparticle Auger Rates in Semiconductor Quantum Dots},
journal = {Science},
volume = {287},
number = {5455},
pages = {1011-1013},
year = {2000},
doi = {10.1126/science.287.5455.1011},
URL = {https://www.science.org/doi/abs/10.1126/science.287.5455.1011},
eprint = {https://www.science.org/doi/pdf/10.1126/science.287.5455.1011}}

@article{carbone2025creation,
    author = {Carbone, Amedeo and Bendixen-Fernex de Mongex, Diane-Pernille and Krasheninnikov, Arkady V. and Wubs, Martijn and Huck, Alexander and Hansen, Thomas W. and Holleitner, Alexander W. and Stenger, Nicolas and Kastl, Christoph},
    title = {Creation and microscopic origins of single-photon emitters in transition-metal dichalcogenides and hexagonal boron nitride},
    journal = {Applied Physics Reviews},
    volume = {12},
    number = {3},
    pages = {031333},
    year = {2025},
    month = {09},
    issn = {1931-9401},
    doi = {10.1063/5.0278132},
    url = {https://doi.org/10.1063/5.0278132},
    eprint = {https://pubs.aip.org/aip/apr/article-pdf/doi/10.1063/5.0278132/20706884/031333_1_5.0278132.pdf},
}

@article{li2019defect,
  author  = {Li, Lesheng and Carter, Emily A.},
  title   = {Defect-Mediated Charge-Carrier Trapping and
             Nonradiative Recombination in {WSe$_2$} Monolayers},
  journal = {Journal of the American Chemical Society},
  volume  = {141},
  number  = {26},
  pages   = {10451--10461},
  year    = {2019},
  doi     = {10.1021/jacs.9b04663}
}

@article{immanuel2022quantum,
    author = {Ghosh Dastidar, Madhura and Thekkooden, Immanuel and Nayak, Pramoda K. and Praveen Bhallamudi, Vidya},
    title = {Quantum emitters and detectors based on 2D van der Waals materials},
    journal = {Nanoscale},
    volume = {14},
    number = {14},
    pages = {5289-5313},
    year = {2022},
    month = {04},
    issn = {2040-3364},
    doi = {10.1039/d1nr08193d},
    url = {https://doi.org/10.1039/d1nr08193d},
    eprint = {https://pubs.rsc.org/nr/article-pdf/14/14/5289/8016481/d1nr08193d.pdf},
}
\bibliographystyle{rsc} 

\clearpage
\newgeometry{
    top=2.5cm,
    bottom=2.5cm,
    left=2.5cm,
    right=2.5cm
}
\onecolumn

\raggedbottom

\pagestyle{plain}



\setcounter{section}{0}
\setcounter{subsection}{0}
\setcounter{figure}{0}
\setcounter{table}{0}
\setcounter{equation}{0}

\renewcommand{\thesection}{S\arabic{section}}
\renewcommand{\thesubsection}{S\arabic{section}.\arabic{subsection}}
\renewcommand{\thefigure}{S\arabic{figure}}
\renewcommand{\thetable}{S\arabic{table}}
\renewcommand{\theequation}{S\arabic{equation}}

\begin{center}

{\LARGE\bfseries Supplementary Information\par}

\vspace{0.6cm}

{\Large\bfseries
Long lived localized defect states in monolayer WSe$_2$: Optical Lifetime distribution and thermal evolution
\par}

\vspace{0.7cm}

{\large
Immanuel Thekkooden$^{a,e,f}$,
Susmita Jana$^{b,c}$,
Mrinal Deka$^{d}$,
B.~R.~K.~Nanda$^{b,c}$,
and V.~Praveen Bhallamudi$^{a,e,f}$
\par}

\vspace{0.6cm}

{\small
$^{a}$Department of Electrical Engineering,
Indian Institute of Technology Madras, Chennai 600036, India\\
$^{b}$Condensed Matter Theory and Computational Lab, Department of Physics,
Indian Institute of Technology Madras, Chennai 600036, India\\
$^{c}$Center of Atomistic Modelling and Materials Design, Department of Physics,
Indian Institute of Technology Madras, Chennai 600036, India\\
$^{d}$Center for 2D Materials Research and Innovation, Department of Physics,
Indian Institute of Technology Madras, Chennai 600036, India\\
$^{e}$Quantum Defects Lab, Department of Physics,
Indian Institute of Technology Madras, Chennai 600036, India\\

$^{f}$Quantum Center of Excellence for Diamond and Emerging Materials (QuCenDiEM) Group, Department of Physics, Indian Institute of Technology Madras, Chennai 600036, India\\
\par}

\vspace{0.4cm}

{\small
$^{*}$Corresponding author:
\href{mailto:praveen.bhallamudi@iitm.ac.in }
{\nolinkurl{praveen.bhallamudi@iitm.ac.in }}
\par}

\end{center}

\vspace{0.8cm}

\hrule

\vspace{1cm}

\startcontents[supplementary]

\begin{center}
{\Large\bfseries Contents\par}
\end{center}

\vspace{0.3cm}

\printcontents[supplementary]{}{1}{
    \setcounter{tocdepth}{2}
}




\section{Confocal Microscopy system}

\subsection{Verifying confocality of system}
\begin{figure}[htbp]
\centering
\includegraphics[width=1.0\textwidth]{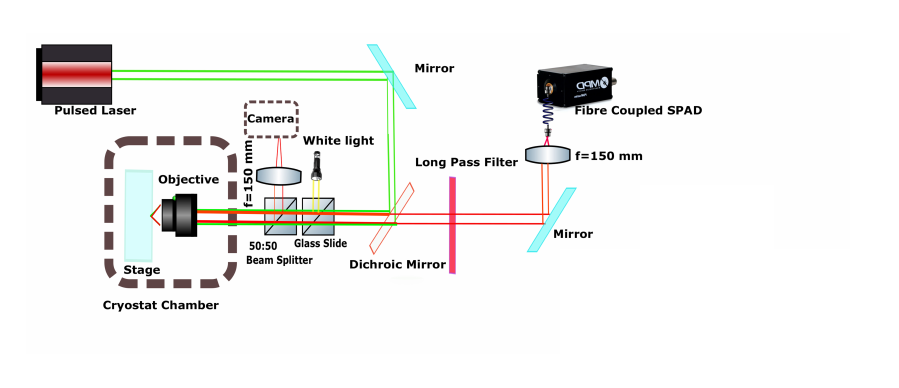}\caption{ Schematic of home-built confocal microscopy system used in our experiment which is intergrated to sample holder and objective in cryostat.}
\label{cryostat}
\end{figure}
A confocal microscope system was built integrating sample holder and objective lens which are inside cryostat chamber as shown in Fig \ref{cryostat}. The system and components are covered in main paper. Here we show the confocality of the system.
The spot size of laser on sample is calculated using equation \cite{meinhart2003theory}
\begin{equation} 
    d_{\infty}=1.22 \lambda_{\text {pump }} \sqrt{\left(\frac{n}{\mathrm{NA}}\right)^{2}-1}
\end{equation}
where $d_{\infty}$ is the size of the infinity corrected spot, n is the refractive index of the medium, NA is the numerical aperture of the objective and $\lambda_{\text pump}$ is the pump wavelength. The spotsize of laser is 447 nm from spotsize equation. At detector side, the PL spot size ($ PL_{d_{\infty}}$) will be increased from $d_{\infty}$ by the factor of magnification of the system with corrected PL wavelength factor.
\begin{equation} 
    PL_{d_{\infty}}=d_{\infty}\frac{\lambda_{\text {PL }}}{\lambda_{\text {pump}}}M
\end{equation}
where M is the magnification of the system (which is calculated to be 52).$PL_{d_{\infty}}$ is calculated as 33.25 um. The fibre core diameter of multimode fibre used at detection end is 50 um. Thus $PL_{d_{\infty}}$ is approximately close to core diameter of multi-mode fibre (which mimics pinhole  in typical confocal system ). We didnt use multimode fibre less than 50 um to ensure , enough signal to noise ratio in our system.

\subsection{Laser spot size measurement}
\begin{figure}[htbp]
\centering
\includegraphics[width=1.0\textwidth]{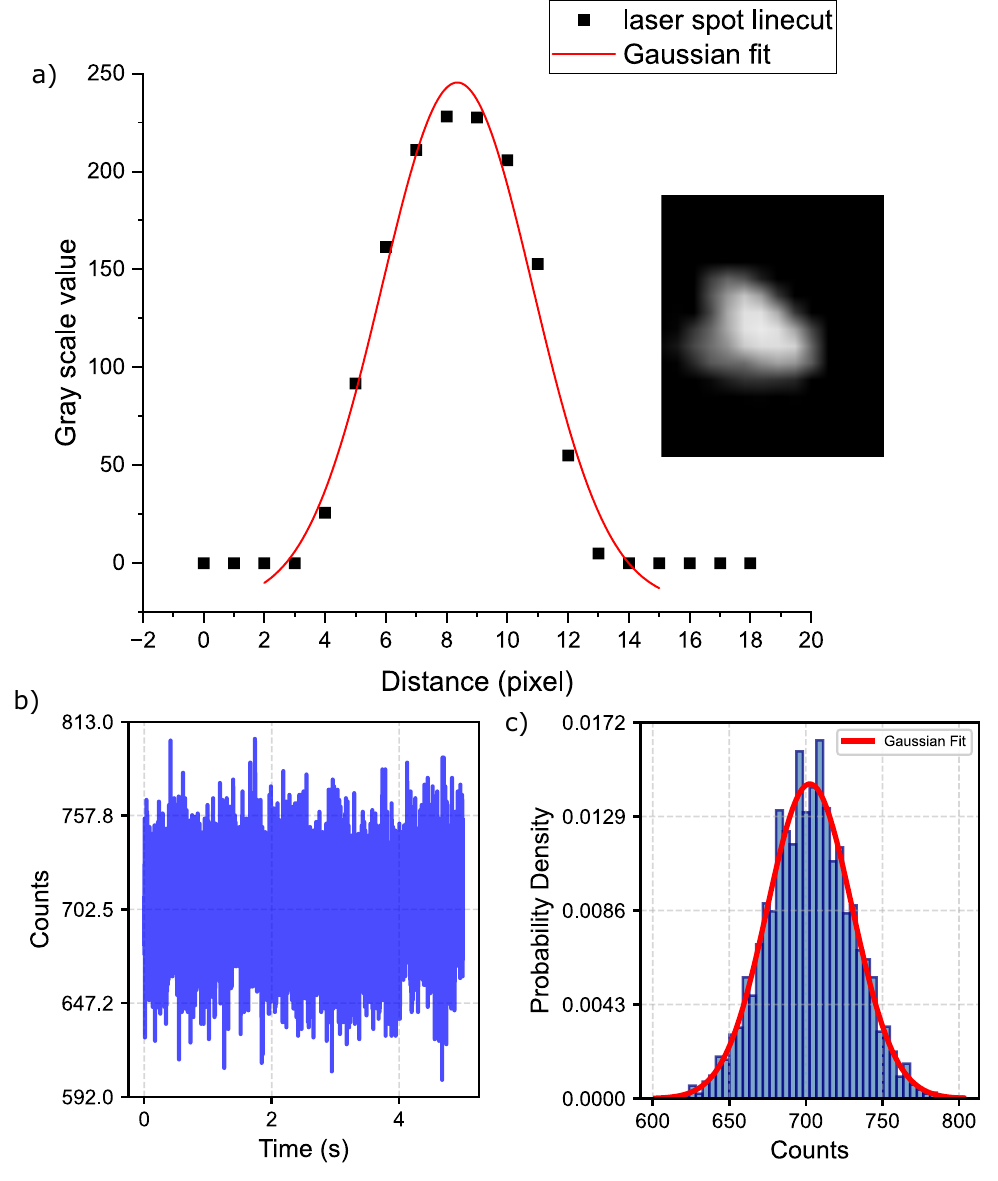}\caption{a) The image of laser spot is captured in camera as shown in inset. The linecut of laser profile is fitted to gaussian b) Time series data of WSe$_2$ PL data is acquired and its mean and variance is calculated. c) The probability density of photon flux distribution is generated from time series data  and is fitted to gaussian distribution.}
\label{spotsize}

\end{figure}

The laser spot size was measured experimentally using a direct imaging technique. To prevent saturation of the camera sensor and ensure a linear response, the laser intensity was attenuated using a set of neutral density filters. The spot was subsequently imaged using a Logitech camera configured with a low exposure setting, as shown in Fig.~\ref{spotsize} a.

The resulting image was analyzed using ImageJ software to extract a line-cut profile of the laser spot intensity, which plots the grayscale value as a function of pixel position. This intensity profile was fitted with a Gaussian function of the form:
\begin{equation}
\label{eq:gaussian}
I(x) = I_0 + \frac{A}{w \sqrt{\pi / 2}} \, \exp\left(-2\left(\frac{x - x_0}{w}\right)^2\right),
\end{equation}
where \(I_0\) is the background offset, \(A\) is a parameter proportional to the area under the curve, \(x_0\) is the center position, and \(w\) is the \(1/e^2\) beam radius parameter. The fit yielded a value of \(w = \qty{4.97}{pixels}\).

The full width at half maximum (FWHM) of the intensity profile is related to the parameter \(w\) by the expression:
\begin{equation}
\label{eq:fwhm}
\text{FWHM} = w \sqrt{2 \ln 2}.
\end{equation}
Substituting the fitted value, the FWHM was calculated to be \qty{4.97}{pixels} \( \times \sqrt{2 \ln 2} \approx \qty{5.85}{pixels}\).

The physical pixel size of the sensor was calibrated by imaging an object of known size under the same optical configuration. Using this calibration factor, we extracted a spotsize of approximately 500 nm.

\subsection{Noise measurement}

To verify that the confocal system's noise performance is limited by the fundamental shot noise of the photon flux, we analyzed the statistical properties of a temporally resolved photon count measurement. In the shot-noise-limited regime, the photon statistics are expected to follow a Poisson distribution, for which the variance is equal to the mean:
\begin{equation}
\label{eq:shot_noise}
\sigma^2 = \mu,
\end{equation}
where \(\sigma^2\) is the variance and \(\mu\) is the mean photon count. For a large number of counts, this Poisson distribution converges to a Gaussian.

The measured time-series data [Fig.~\ref{spotsize}(b)] yielded a mean
count of \(\mu = 702.56\) counts per acquisition interval and a variance
of \(\sigma^{2} = 767.88~\mathrm{counts}^{2}\). The near equivalence of the mean and variance (\(\mu \approx \sigma^2\)) provides strong evidence for shot-noise dominance. Additionally, the probability density distribution of the photon counts conforms to a Gaussian profile centered at the mean (Fig.~\ref{spotsize} c), consistent with the expected behavior of a shot-noise process at high photon fluxes. These results collectively confirm that the system is operating in the shot-noise-limited regime.

\section{\texorpdfstring
{Raman and PL data of WSe$_2$}
{Raman and PL data of WSe2}}

\begin{figure}[htbp]
\centering
\includegraphics[width=1.0\textwidth]{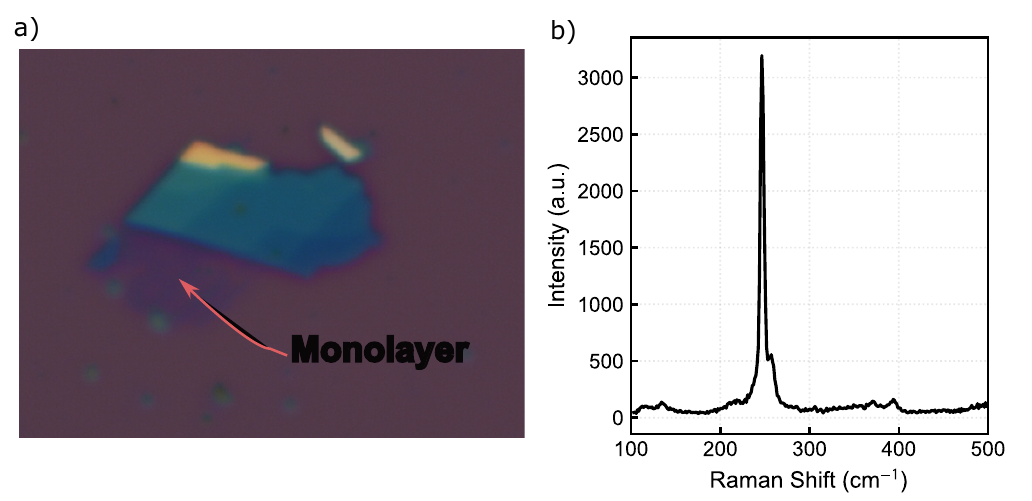}\caption{a) Optical microscope image of WSe$_2$ flake. The monolayer region is marked using arrows. b) Raman spectra of WSe$_2$ flake in room temperature }
\label{Raman}

\end{figure}
\begin{figure}[htbp]
\centering
\includegraphics[width=0.8\textwidth]{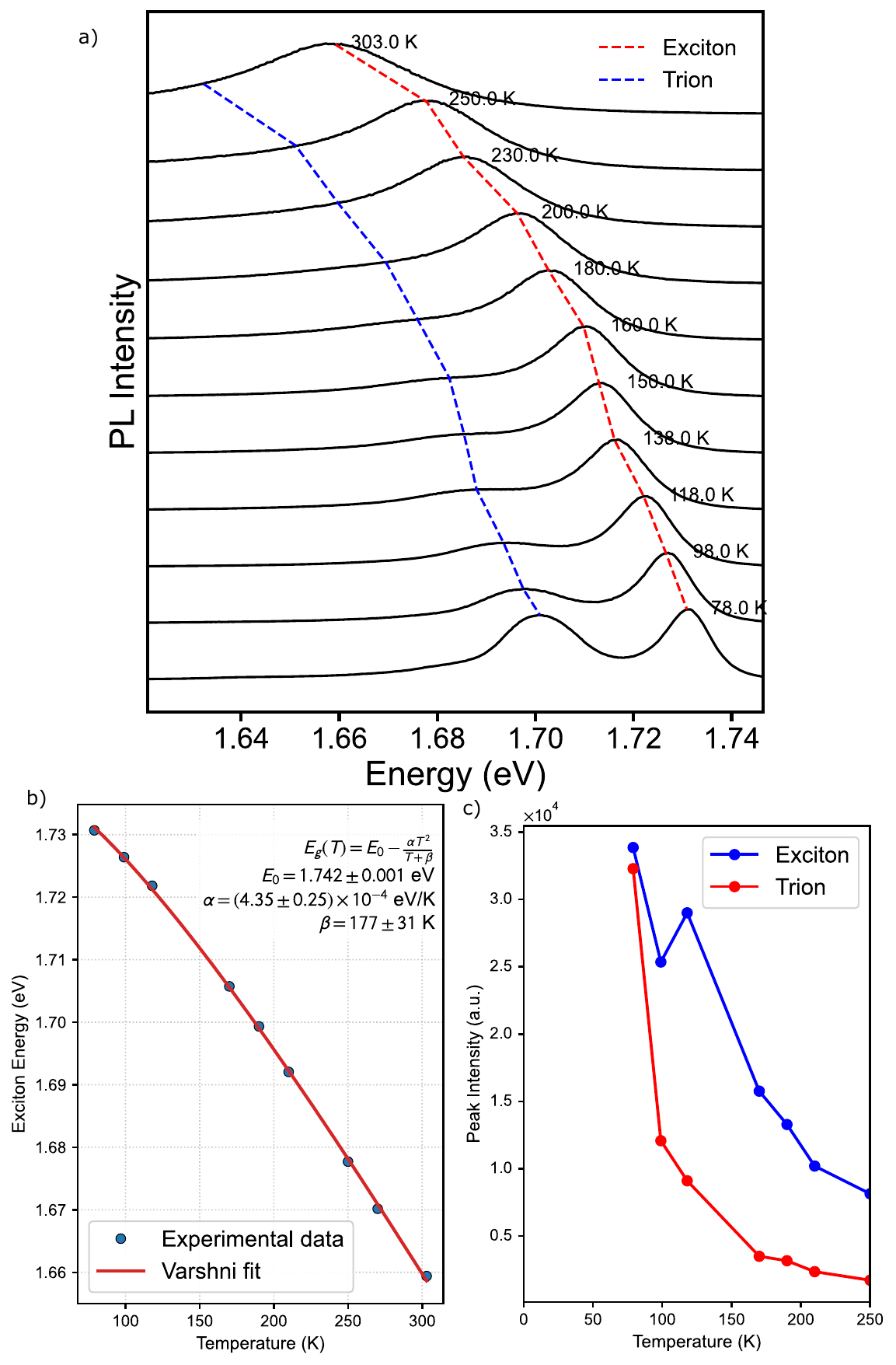}\caption{a) PL data of WSe$_2$ at higher temperature b) The exciton position in PL data is fitted to Varshini model to understand band-gap variation in WSe$_2$ as function of temperature. c)  The intensity variation of exciton and trion peak with respect to temperature. }
\label{PL}

\end{figure}
\subsection{\texorpdfstring
{Optical Image of WSe$_2$ and Raman Data of WSe$_2$}
{Optical Image of WSe2 and Raman Data of WSe2}}
The optical microscope image of 
WSe$_2$ is shown in  Fig.~\ref{Raman} a. 

Raman spectroscopy was used to characterize the structural properties of the  WSe$_2$ sample at room temperature. The measured spectrum is presented in Fig.~\ref{Raman} b. Two prominent first-order Raman modes are observed: one mode at \qty{246.5}{\per\centi\meter} and  other mode at \qty{257.2}{\per\centi\meter}. The frequency separation between these two characteristic peaks is \qty{10.7}{\per\centi\meter}\cite{lu2015atomic}.

This measured peak separation is a key indicator of layer thickness in transition metal dichalcogenides (TMDCs). For WSe$_2$, the value of \qty{10.7}{\per\centi\meter} is consistent with reported values for a monolayer flake \cite{tonndorf2013photoluminescence,sahin2013anomalous}.

Furthermore, the spectrum exhibits a clear absence of the layer-breathing mode (LBM), which is typically observed near \qty{308}{\per\centi\meter} in bilayer and few-layer WSe$_2$ \cite{Zhao2013}. The lack of this vibrational signature provides definitive confirmation that the measured region is exclusively monolayer, with no detectable bilayer or few-layer domains.

\subsection{\texorpdfstring
{PL data of WSe$_2$ }
{PL data of WSe2}}

Temperature-dependent photoluminescence (PL) spectroscopy was performed on a WSe$_2$ sample across a temperature range from room temperature down to \qty{78}{K}, as shown in Fig.~\ref{PL} a     . At room temperature, the PL spectrum features a broad emission band where the neutral exciton (X$^0$) and charged trion (X$^-$) contributions are not resolvable without spectral deconvolution. A double Lorentzian fit is required to separate these components accurately.

As the temperature decreases, the overall PL intensity increases and the spectral features sharpen. Concurrently, the trion emission intensity grows relative to the exciton, leading to a clear distinction between the two peaks at lower temperatures, as illustrated in Fig.~\ref{PL} a. Data acquired below \qty{78}{K} are presented and discussed in the main text.

\subsection{Exciton energy shift and intensity variation as a function of temperature}
\label{ssec:varshni}

The energy of the neutral-exciton PL peak was tracked as a function of
temperature. The observed redshift with increasing temperature is
consistent with the temperature-dependent renormalization of the
electronic band structure arising from lattice expansion and
electron--phonon interactions\cite{guerra2024temperature}. The measured
exciton transition energy was fitted using the empirical Varshni
relation,

\begin{equation}
\label{eq:varshni}
E_X(T)
=
E_X(0)
-
\frac{\alpha T^2}{T+\beta},
\end{equation}

where $E_X(T)$ is the neutral-exciton transition energy at temperature
$T$, $E_X(0)$ is its extrapolated value at 0 K, $\alpha$ is the
temperature coefficient, and $\beta$ is an empirical parameter related
to a characteristic phonon temperature.

As shown in Fig.~\ref{PL}(b), the experimental data are well described
by Eq.~(\ref{eq:varshni}). The fit yields

\begin{equation}
E_X(0)=1.742\pm0.001~\mathrm{eV},
\end{equation}

\begin{equation}
\alpha=(4.35\pm0.25)\times10^{-4}~\mathrm{eV\,K^{-1}},
\end{equation}

and

\begin{equation}
\beta=177.5\pm31.4~\mathrm{K}.
\end{equation}

The extrapolated value of $E_X(0)$ is in close agreement with the
neutral-exciton transition energy measured at 4 K.

It is important to distinguish the exciton transition energy measured
by PL from the quasiparticle band gap. These quantities are related by

\begin{equation}
\label{eq:binding}
E_X(T)=E_g(T)-E_b(T),
\end{equation}

where $E_g(T)$ is the quasiparticle band gap and $E_b(T)$ is the exciton
binding energy. Consequently, the Varshni fit directly describes the
temperature dependence of the excitonic optical transition. Its
variation reflects the underlying band-gap renormalization, provided
that changes in the exciton binding energy over the investigated
temperature range are comparatively small.

The integrated intensities of the neutral-exciton and trion peaks,
obtained from the areas of the fitted Lorentzian components, are shown
as a function of temperature in Fig.~\ref{PL}(c). Both emission
features decrease in intensity with increasing temperature. This
thermal quenching indicates the increasing importance of competing
nonradiative recombination pathways, thermal dissociation, and carrier
redistribution among the available excitonic states. 

\section{Lifetime}
Lifetime measurements were conducted for each quasiparticles that is identified from PL. Also Instrument Response Function (IRF) and exciton lifetime at room temperature is studied.  

\subsection{IRF }
\begin{figure}[htbp]
    \centering
    \includegraphics[width=1.0\textwidth]{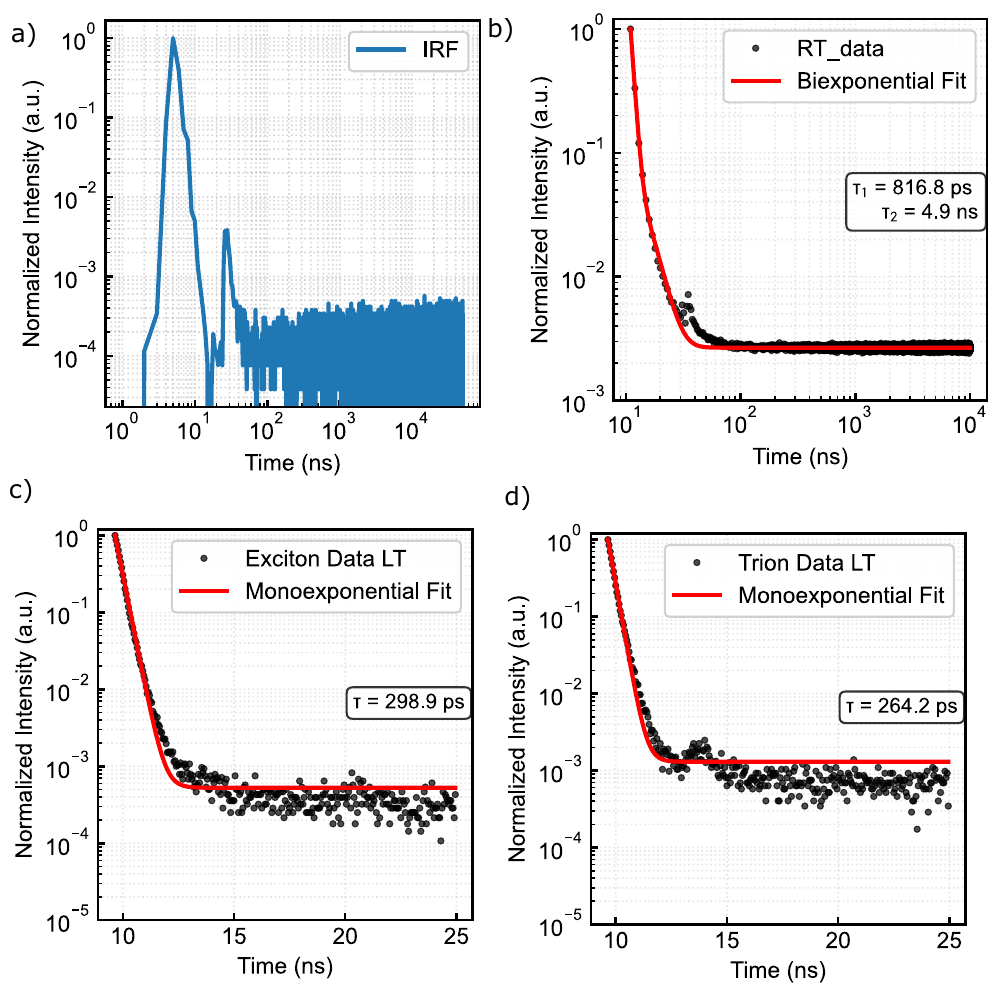}
    \caption{ a) Instrument response function. Solid line represents guide to eye b) Room temperature lifetime data from PL peak of WSe$_2$ at room temperature c) Life-time of exciton peak at 4 K fitted to mono exponential decay model. d) Life-time of trion peak at 4 K fitted to mono exponential decay model }
    \label{IRF}
\end{figure}
The ideal excitation source for time-resolved photoluminescence (TRPL) measurements is an impulse function (a Dirac delta pulse) characterized by infinitesimally fast rise and fall times. In practice, however, such an ideal pulse is physically unattainable. The measured temporal profile is inevitably broadened by the finite response time of every component within the experimental apparatus.

The primary sources of temporal broadening include the intrinsic pulse width of the laser, the chromatic dispersion of the optical system, the finite response time and temporal jitter of the detector, and the timing limitations of the associated electronics. The cumulative effect of these individual broadening mechanisms is described by the Instrument Response Function (IRF), which is mathematically defined as the convolution of the response functions of all system components:

\begin{equation}
\text{IRF}(t) = f_{\text{laser}}(t) \ast f_{\text{optics}}(t) \ast f_{\text{detector}}(t) \ast f_{\text{electronics}}(t)
\end{equation}

where the symbol $\ast$ denotes the convolution operation.

As a result, the signal $S_{\text{measured}}(t)$ acquired by the instrument is not the true photoluminescence decay $PL(t)$ of the sample, but rather the convolution of this intrinsic decay with the IRF:

\begin{equation}
S_{\text{measured}}(t) = PL(t) \ast \text{IRF}(t)
\end{equation}

Therefore, to extract the genuine carrier dynamics and the intrinsic lifetime from the sample, the instrumental contribution must be accounted for. This necessitates a numerical deconvolution of the IRF from the measured signal to recover the actual PL response, $PL(t)$

The instrument response function (IRF) of the time-resolved photoluminescence system was characterized by directing the scattered laser pulse directly onto the detector. To prevent detector saturation and ensure a linear response, the laser intensity was significantly attenuated using multiple neutral-density filters.The IRF measured is shown in \ref{IRF} a. A Gaussian fit to the IRF data yielded a full width at half maximum (FWHM) of approximately \textbf{1 ns}, which defines the temporal resolution limit of our setup. Consequently, the accuracy of lifetime components shorter than  to this value is inherently limited. In the main text, the shortest lifetime component ($\tau_0$) falls within the uncertainty margin of the IRF width, and its value should therefore be interpreted with caution. In contrast, the longer lifetime components ($\tau_1$ and $\tau_2$) are significantly longer than the IRF FWHM, ensuring that their extracted values are reliable and largely unaffected by the system's temporal resolution. For this reason, a numerical deconvolution procedure was deemed unnecessary for the analysis of ($\tau_1$ and $\tau_2$) presented in the main text. 

A secondary, low-intensity peak is observed at approximately 22 ns far from main peak in the IRF. This feature is attributed to the known afterpulsing effect in the Excelitas single-photon avalanche diode (SPAD) detector, where trapped charge carriers are released after a characteristic delay corresponding to the detector's dead time. This ~22 ns signature is consistent with previously reported behaviour for this family of detectors\cite{ziarkash2018comparative} . 
\subsection{Brute force iterative convolution}
\begin{figure}[htbp]
\centering
\includegraphics[height=9cm]{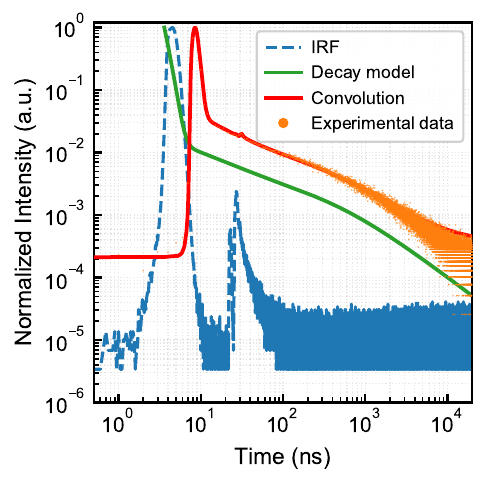}
\caption{Brute-force convolusion of IRF data with physical model. Result of convolusion matches with obtained life-time data}
\label{bruteforce}
\end{figure}
To verify that the proposed decay model accurately reproduces the experimentally measured time-resolved photoluminescence (TRPL), a brute-force convolution approach was employed. The measured instrument response function (IRF), obtained from direct laser reflection, was numerically convolved with the intrinsic decay model using a fast Fourier transform (FFT). This procedure accounts for the finite temporal response of the detection system without introducing any analytical approximations.

The intrinsic decay was modeled as

\begin{equation}
I(t)=A\exp\!\left(-\frac{t}{\tau_{0}}\right)
+\frac{m}{t+\tau_{1}}
+\frac{e}{t+\tau_{2}},
\label{eq:bruteforce_model}
\end{equation}

where the exponential term represents the fast radiative recombination channel, while the two inverse-time terms describe two independent distributions of defect detrapping times. The first power-law component is characterized by the timescale $\tau_{1}$ and the second by $\tau_{2}$.

The calculated decay was convolved with the experimentally measured IRF according to

\begin{equation}
I_{\mathrm{conv}}(t)=
\int_{0}^{\infty}
IRF(t-t')
\,I(t')\,dt',
\label{eq:convolution}
\end{equation}

which was evaluated numerically using FFT convolution. The resulting convolved signal was directly compared with the experimental TRPL data without any additional fitting parameters.

As shown in Fig.~\ref{bruteforce}, the simulated decay accurately reproduces the measured lifetime over more than five orders of magnitude in intensity. The excellent agreement demonstrates that the extracted amplitudes and lifetime are very well reliable and the influence of IRF on it is much less. This numerical validation independently confirms the physical model used throughout the main text.
\subsection{ Noise and detector dark-count analysis}
\label{ssec:dark_count_analysis}
\begin{figure}[htbp]
\centering
\includegraphics[height=9cm]{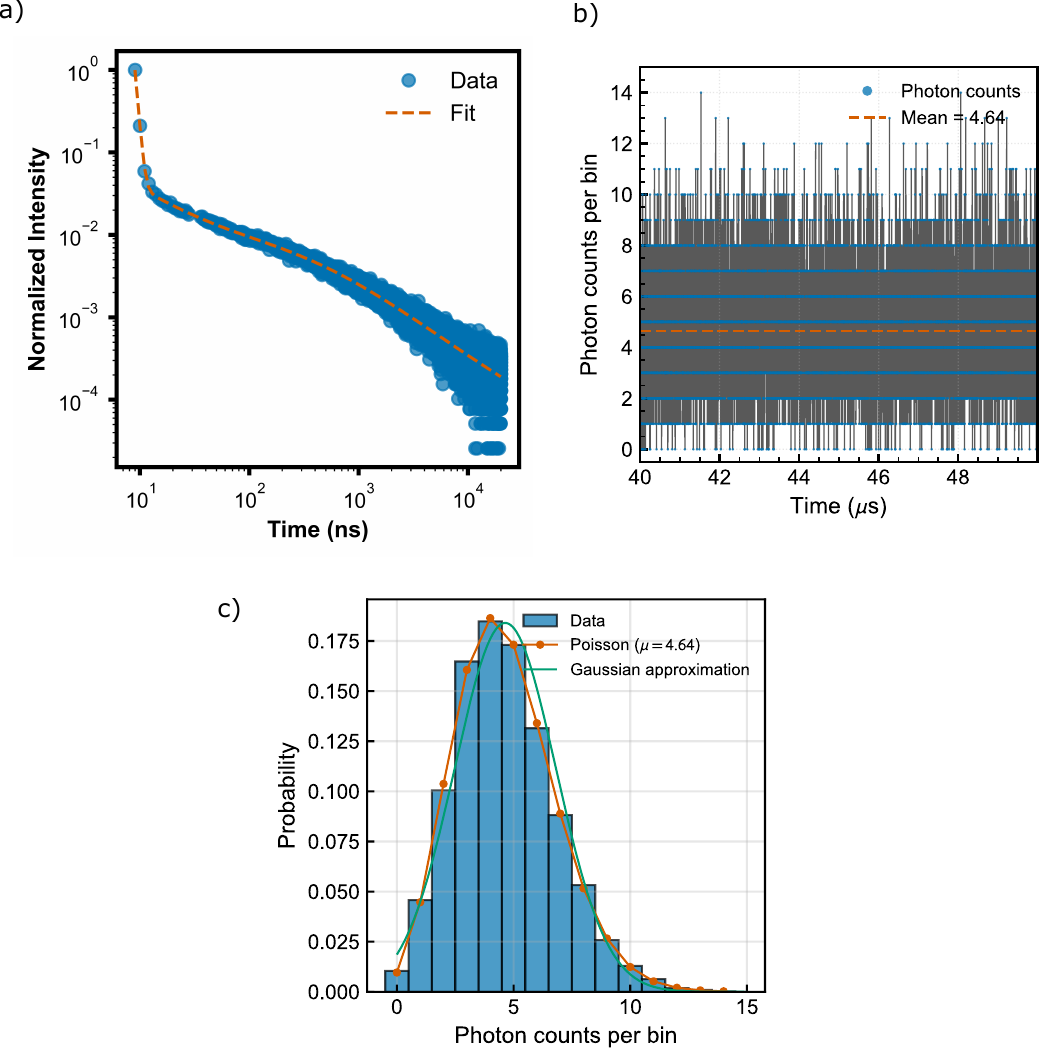}
\caption{a) Long-lifetime data which was collected for $50~\mu\mathrm{s}$. It is seen that decay ends after around $30~\mu\mathrm{s}$.b) From remaining $20~\mu\mathrm{s}$ last $10~\mu\mathrm{s}$ data is zoomed in c) Histogram of photon number distribution of b) is plotted and fitted to a gaussian function }
\label{last1000}
\end{figure}

The complete time-correlated single-photon counting trace contains
50,000 temporal bins with a bin width of 1 ns, corresponding to a total
measurement window of

\begin{equation}
T_{\mathrm{window}}
=
50{,}000\times 1~\mathrm{ns}
=
50~\mu\mathrm{s}.
\end{equation}

The histogram was accumulated over a total acquisition time of
\(T_{\mathrm{acq}}=300~\mathrm{s}\). To characterize the noise floor at
late delay times, where the defect-related PL signal has decayed close
to the background level, the final 10,000 bins of the histogram were
analysed. These points span the final \(10~\mu\mathrm{s}\) of the
measurement window, as shown in Fig.~\ref{last1000}(b).

The late-time count distribution has a mean count of

\begin{equation}
\overline{n}=4.6437~\mathrm{counts~bin^{-1}},
\end{equation}

and a variance of

\begin{equation}
\sigma^{2}=4.6996~\mathrm{counts^{2}~bin^{-1}}.
\end{equation}

The corresponding variance-to-mean ratio is

\begin{equation}
\frac{\sigma^{2}}{\overline{n}}
=
\frac{4.6996}{4.6437}
=
1.012,
\end{equation}

which is close to the value of unity expected for Poisson-distributed
photon-counting noise. A chi-square goodness-of-fit test also yields
\(p=0.8669\), indicating no statistically significant deviation from a
Poisson distribution [Fig.~\ref{last1000}(c)]. The late-time
signal is therefore consistent with shot-noise-limited photon counting.

Assuming that the mean late-time background is uniform throughout the
50-\(\mu\mathrm{s}\) histogram window, the total number of background
counts accumulated during the 300-s acquisition is estimated as

\begin{equation}
N_{\mathrm{bg}}
=
\overline{n}N_{\mathrm{bin}}
=
4.6437\times50{,}000
=
232{,}185,
\end{equation}

where \(N_{\mathrm{bin}}=50{,}000\) is the total number of temporal bins.
The corresponding background count rate is therefore

\begin{equation}
R_{\mathrm{bg}}
=
\frac{N_{\mathrm{bg}}}{T_{\mathrm{acq}}}
=
\frac{232{,}185}{300~\mathrm{s}}
=
773.95~\mathrm{s^{-1}}.
\end{equation}

Thus, the estimated late-time background rate is approximately

\begin{equation}
\boxed{R_{\mathrm{bg}}\approx774~\mathrm{counts~s^{-1}}.}
\end{equation}

The same result can be obtained directly from the final 10,000 bins.
These bins contain, on average,

\begin{equation}
4.6437\times10{,}000=46{,}437
\end{equation}

counts over 300 s, corresponding to \(154.8~\mathrm{s^{-1}}\) within
the selected 10-\(\mu\mathrm{s}\) interval. Since this interval
represents one-fifth of the complete 50-\(\mu\mathrm{s}\) histogram
window, the equivalent full-window rate is

\begin{equation}
5\times154.8~\mathrm{s^{-1}}
=
774~\mathrm{s^{-1}}.
\end{equation}

The magnitude and Poissonian statistics of this late-time count floor
are consistent with the detector dark-count background. Laser blocked measurement independently yielded a detector dark count rate of 700 counts per second.
    
\subsection{Lifetime of different quasi-particles}

At room temperature, as established previously, the photoluminescence (PL) spectrum is dominated by excitonic emission, with the trion contribution manifesting as a low-energy tail. Time-resolved PL measurements of the excitonic peak (Fig.~\ref{IRF} b) were performed to probe its recombination dynamics. The decay curve was well-described by a bi-exponential function, yielding characteristic decay constants of $\tau_1 = \SI{816}{\pico\second}$ and a minor, longer component of $\tau_2 = \SI{4.9}{\nano\second}$ with low amplitude\cite{Mouri_2014}. Such a biexponential decay profile, indicative of multiple recombination pathways particularly involving defect trapping, has been previously reported for excitons in monolayer transition metal dichalcogenides \cite{palummo2015exciton,Mouri_2014}.

To isolate the intrinsic radiative properties, time-resolved PL was also measured at $\SI{4}{\kelvin}$ using a $\SI{710}{\nano\meter}$ bandpass filter ($\SI{10}{\nano\meter}$ FWHM) to selectively probe the exciton emission. At this temperature, the decay kinetics simplified and were adequately fit by a single-exponential function, yielding a lifetime of $\tau_X = \SI{298}{\pico\second}$ (Fig.~\ref{IRF} c). Although this value is near the temporal resolution limit of our system (IRF FWHM $\approx \SI{1}{\nano\second}$), it establishes an upper bound for the exciton lifetime, confirming it to be $\leq \SI{298}{\pico\second}$. This finding is consistent with the range of ultrafast exciton lifetimes ($\sim\SI{5}{\pico\second}$ to $\SI{200}{\pico\second}$) reported in the literature.

Similarly, the trion dynamics were probed at $\SI{4}{\kelvin}$ using a $\SI{720}{\nano\meter}$ bandpass filter as shown in Fig ~\ref{IRF}  d    . The trion decay was also monoexponential, with a fitted lifetime of $\tau_T = \SI{264}{\pico\second}$. Analogous to the exciton case, the measured value represents an upper limit, with the actual trion lifetime likely being less than or equal to $\SI{264}{\pico\second}$.

\subsection{Historical Background of the \texorpdfstring{$1/t$}{1/t} Decay }

A non-exponential decay of the form

\begin{equation}
I(t)\propto \frac{1}{t+t_{0}}
\label{eq:1tovert}
\end{equation}

has been widely reported in systems where the relaxation dynamics are governed by either
bimolecular interactions or a broad distribution of recombination rates. 

In semiconductor optics, a similar temporal dependence naturally emerges from
exciton-exciton annihilation (EEA). The exciton population density \(n(t)\) undergoing
EEA obeys the rate equation

\begin{equation}
\frac{dn}{dt}
=
-\gamma n^{2},
\label{eq:eea_rate}
\end{equation}

where \(\gamma\) is the annihilation coefficient. Solving Eq.~(\ref{eq:eea_rate}) gives

\begin{equation}
n(t)
=
\frac{n_{0}}
{1+\gamma n_{0}t}
=
\frac{1/\gamma}
{t+t_{0}},
\qquad
t_{0}=\frac{1}{\gamma n_{0}}.
\label{eq:eea_solution}
\end{equation}

Thus, at long times the population follows the universal asymptotic behaviour

\begin{equation}
n(t)\propto \frac{1}{t},
\end{equation}

which has been observed in numerous low-dimensional systems including quantum
wells\cite{klimovquantization2020}, carbon nanotubes\cite{Mainfluenceof2015}, transition-metal dichalcogenides, perovskites, and colloidal
quantum dots\cite{Pengbright2020}, where exciton diffusion and many-body interactions enhance the
probability of exciton-exciton encounters.


The observation of a $1/t$-like decay does not uniquely identify
exciton--exciton annihilation. Similar algebraic relaxation may arise
from a broad distribution of recombination, trapping, or switching
times associated with energetic and spatial disorder. Power-law and broadly distributed relaxation kinetics have also been
reported in several physically distinct material systems and can
originate from mechanisms other than bimolecular annihilation.
Representative examples include:

\begin{itemize}
    \item \textbf{Distributed photoluminescence decay in halide
    perovskites.} Yuan \textit{et al.} reported transient
    photoluminescence extending from tens of nanoseconds to hundreds of
    microseconds in triple-cation perovskite films
    \cite{yuan2024shallow}. Rather than being characterized by a single
    lifetime, the decay exhibited continuously varying instantaneous
    lifetimes and was quantitatively explained by carrier trapping,
    thermal emission, and recombination through a high density of
    shallow defect states. This study directly illustrates how
    defect-mediated carrier kinetics can produce long-lived,
    non-exponential photoluminescence for which the assignment of a
    single discrete lifetime is not physically meaningful.

    \item \textbf{Long-time quantum decay in inverse-square
    potentials.} Martorell \textit{et al.} showed that the survival
    probability of a particle escaping from a trapping potential can
    cross over from exponential to power-law relaxation at long times.
    For a long-range inverse-square potential, the power-law exponent
    can vary continuously with the strength of the potential tail
    \cite{PhysRevA.77.042719}. This result illustrates that algebraic
    relaxation can arise intrinsically from the spectral and spatial
    properties of a confined quantum system.

    \item \textbf{Fluorescence intermittency in CdSe quantum dots.}
    Kuno \textit{et al.} observed inverse-power-law distributions of
    the bright and dark residence times of individual CdSe/ZnS quantum
    dots over a broad range of timescales
    \cite{kuno2000nonexponential}. The behaviour was interpreted as a
    signature of distributed kinetics arising, for example, from a
    distribution of trap depths or tunnelling distances between the
    quantum-dot core and interfacial states.

    \item \textbf{Power-law blinking determined by autocorrelation
    analysis.} Houel \textit{et al.} developed a
    threshold-independent autocorrelation method for extracting the
    bright- and dark-state power-law exponents of individual CdSe/CdS
    quantum dots\cite{houel2015autocorrelation}. Their analysis showed
    that power-law blinking statistics can be recovered even from
    shot-noise-dominated intensity traces, although the work primarily
    establishes an analysis methodology rather than a unique
    microscopic origin of the kinetics.

    \item \textbf{Statistical decay of isolated molecular clusters.}
    Andersen \textit{et al.} investigated the cooling dynamics of
    isolated clusters possessing a broad distribution of internal
    energies and microcanonical temperatures
    \cite{andersen2003temperature}. The ensemble-averaged decay rate was
    found to exhibit an approximately $1/t$ dependence over an
    intermediate time range, before crossing over to an almost
    exponential decay when radiative cooling became important. This
    work demonstrates that a $1/t$-like relaxation can emerge from the
    statistical averaging of many systems having a broad distribution
    of decay rates, without invoking bimolecular annihilation.
\end{itemize}

These examples demonstrate that power-law or strongly non-exponential
relaxation is not specific to a single material class or microscopic
interaction. In the present system, the spatial localization of the
defect-bound excitons, together with the absence of systematic
fluence-induced shortening of the long-lived components, favours an
interpretation based on broadly distributed defect-related
recombination rates rather than an exciton--exciton-annihilation-
dominated process.

The more general power-law relaxation curve with a stretching exponent $\beta$ can be interpreted as emerging from a distribution of relaxation rates across various timescales~\cite{widder1941laplace}. Mathematically, this can be expressed as:

\begin{equation}
\frac{1}{(1 + t/\tau_0)^\beta} = \int_0^\infty P(s, \beta) e^{-st/\tau_0} \, ds,
\label{eq:stretched_power_law_distribution}
\end{equation}

where $\tau_0$ is the characteristic decay time, $s = \tau_0/\tau = \Gamma/\Gamma_0$ is the dimensionless relaxation rate parameter, $\Gamma_0 = 1/\tau_0$ is the characteristic rate obtained from fits to the stretched power-law model, and the local relaxation rate $\Gamma = 1/\tau$ is a property of a particular defect center contributing to the total signal.

The distribution of local rates $\Gamma$ is expressed with the stretching exponent $\beta$ as:

\begin{equation}
P(s, \beta) = \frac{s^{\beta-1} e^{-s}}{\Gamma(\beta)},
\label{eq:rate_distribution}
\end{equation}

where $\Gamma(\beta)$ is the gamma function. For $\beta \to \infty$, the distribution approaches a Dirac delta distribution centered at $s = 1$:

\begin{equation}
\lim_{\beta \to \infty} P(s, \beta) = \delta(s-1),
\label{eq:dirac_delta_limit}
\end{equation}

corresponding to a single well-defined value of $1/\tau_0$. For finite $\beta$, the distribution becomes broadened and skewed, with the relaxation rate $\Gamma_{\text{max}} < \Gamma_0$ at which $P(s, \beta)$ has a maximum, and a long tail towards $\Gamma > \Gamma_0$.

For the specific case of $\beta = 1$, the distribution reduces to a simple exponential $P(s) = e^{-s}$, corresponding to the standard $1/t$ decay. For $\beta < 1$, the distribution diverges at $s = 0$, indicating a broad distribution dominated by very slow rates (long lifetimes). For $\beta > 1$, the distribution peaks at $s_{\text{max}} = \beta - 1$, corresponding to a most probable rate of $\Gamma_{\text{max}} = (\beta - 1)\Gamma_0$.

Additional examples of non-exponential relaxation other than $1/t$ arising from
distributions of microscopic environments, trapping times, or
donor--acceptor separations include:

\begin{itemize}

    \item \textbf{Energy transfer among randomly distributed centres.}
    Inokuti and Hirayama showed that luminescence quenching by energy
    transfer from an excited donor to randomly distributed acceptors
    produces a non-exponential decay. The decay reflects the spatial
    distribution of donor--acceptor separations and the strong
    distance dependence of the transfer rate
    \cite{inokuti1965influence}.

    \item \textbf{Localized excitons in disordered quantum wells.}
    Photoluminescence from localized states in
    InGaAsN/GaAs quantum wells was found to follow a stretched
    exponential decay at low temperature. This behaviour was attributed
    to a distribution of localized environments generated by random
    alloy-potential fluctuations
    \cite{takagishi2007stretched}.

    \item \textbf{Donor--acceptor recombination in ZnS phosphors.}
    Strongly non-exponential photoluminescence was reported in
    ZnS:Cu,Al phosphors, with the decay evolving between stretched
    exponential and power-law-like forms depending on the excitation
    conditions. The behaviour was described using spatially correlated
    donor and acceptor states possessing a broad distribution of
    recombination times
    \cite{martin2008onedimensional}.

    \item \textbf{Defect-related donor--acceptor-pair emission in GaN.}
    At low temperatures, defect-related yellow and red luminescence
    bands in GaN exhibit non-exponential decays that are close to
    power-law behaviour. These kinetics were associated with
    donor--acceptor-pair transitions, for which the recombination rate
    varies strongly with the spatial separation of the participating
    centres
    \cite{reshchikov2018two}.

    \item \textbf{Dispersive trapping in amorphous semiconductors.}
    Scher and Montroll developed a continuous-time random-walk
    description of transport in disordered solids in which a broad
    distribution of trapping and waiting times produces algebraic,
    rather than exponential, relaxation
    \cite{scher1975anomalous}. Although this model describes carrier
    transport rather than photoluminescence directly, it provides a
    general theoretical basis for power-law kinetics arising from
    energetic and spatial disorder.

\end{itemize}

\subsection{Lifetime distribution corresponding to the power-law decay}

The long-lived components of the defect emission were described by power-law decay terms of the form

\begin{equation}
I(t)=\frac{C}{t+\tau_0},
\end{equation}

where $C$ is the amplitude and $\tau_0$ is the characteristic time obtained from the fit. Such a decay can be interpreted as the superposition of a continuum of exponential decays,

\begin{equation}
\frac{C}{t+\tau_0}
=
\int_{0}^{\infty}
P_{\Gamma}(\Gamma)
e^{-\Gamma t}\,
d\Gamma,
\end{equation}

where $\Gamma$ denotes the local recombination rate. Using the Laplace transform identity,

\begin{equation}
P_{\Gamma}(\Gamma)
=
\frac{C}{\Gamma_0}
\exp\!\left(-\frac{\Gamma}{\Gamma_0}\right),
\qquad
\Gamma_0=\frac{1}{\tau_0}.
\end{equation}

Expressing the distribution in terms of lifetime
$\tau=1/\Gamma$ gives

\begin{equation}
P_{\tau}(\tau)
=
C\,
\frac{\tau_0}{\tau^{2}}
\exp\!\left(-\frac{\tau_0}{\tau}\right),
\label{eq:lifetime_distribution}
\end{equation}

which represents the probability density of local defect lifetimes contributing to the measured decay. The distribution possesses a broad $1/\tau^{2}$ tail at long lifetimes, reflecting the absence of a single characteristic trapping time and indicating a continuum of localized defect environments.

\begin{figure}
\centering
\includegraphics[height=9cm]{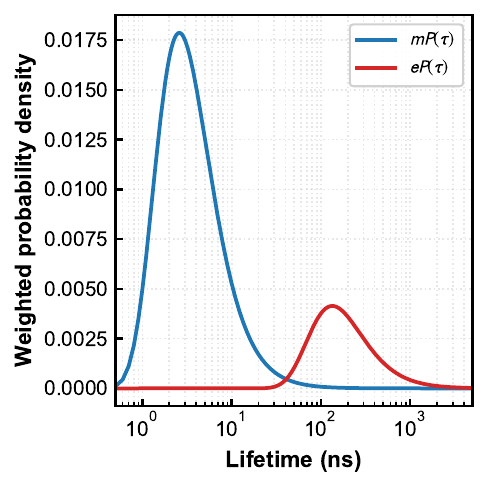}
\caption{Amplitude-weighted lifetime distributions obtained from the two power-law decay channels. The distributions were calculated using Eq.~(\ref{eq:lifetime_distribution}) with the fitted values of $\tau_1$, $\tau_2$, $m$, and $e$. The maxima occur at $\tau=\tau_1/2$ and $\tau=\tau_2/2$, respectively.}
\label{fig:lifetime_distribution}
\end{figure}

For the present work, the two power-law components,

\begin{equation}
\frac{m}{t+\tau_1}
\qquad\text{and}\qquad
\frac{e}{t+\tau_2},
\end{equation}

were independently converted into the corresponding lifetime distributions using Eq.~(\ref{eq:lifetime_distribution}) and weighted by their fitted amplitudes (characteristic lifetimes, $m$ and $e$ used from Fig 2 d of main paper). The resulting distributions are shown in Fig.~\ref{fig:lifetime_distribution}. The short-lived channel peaks near $\tau_1/2$, whereas the long-lived channel peaks near $\tau_2/2$, demonstrating the existence of two distinct detrapping populations spanning a broad range of defect lifetimes.

\subsection{Exciton concentration under pulsed excitation}
The time-averaged carrier density was calculated using:

\begin{equation}
n_0 = \frac{\eta A (1-R) P_{\text{avg}}}{E_{\text{ph}} A_{\text{spot}} f},
\end{equation}

with the following parameters: peak power = 4~mW, pulse width = 5~ns, pulse period = 50~$\mu$s (duty cycle = 0.010\%), average power ($P_{\text{avg}}$) = 0.4~$\mu$W, excitation spot ($A_{\text{spot}}$) = $0.5 \times 0.5$~$\mu$m$^2$, absorbance(A) = 3\%, Reflectance (R)=20\%, photon energy($E_{\text{ph}}$) = 2.36~eV, repetition rate = 20~kHz, and collection efficiency($\eta$) = 0.6. This yields $n_0 = 3.05 \times 10^{13}$~cm$^{-2}$.

\subsection{Effect of excitation fluence on the defect-emission dynamics}

\begin{figure*}
    \centering
    \includegraphics[height=9 cm]{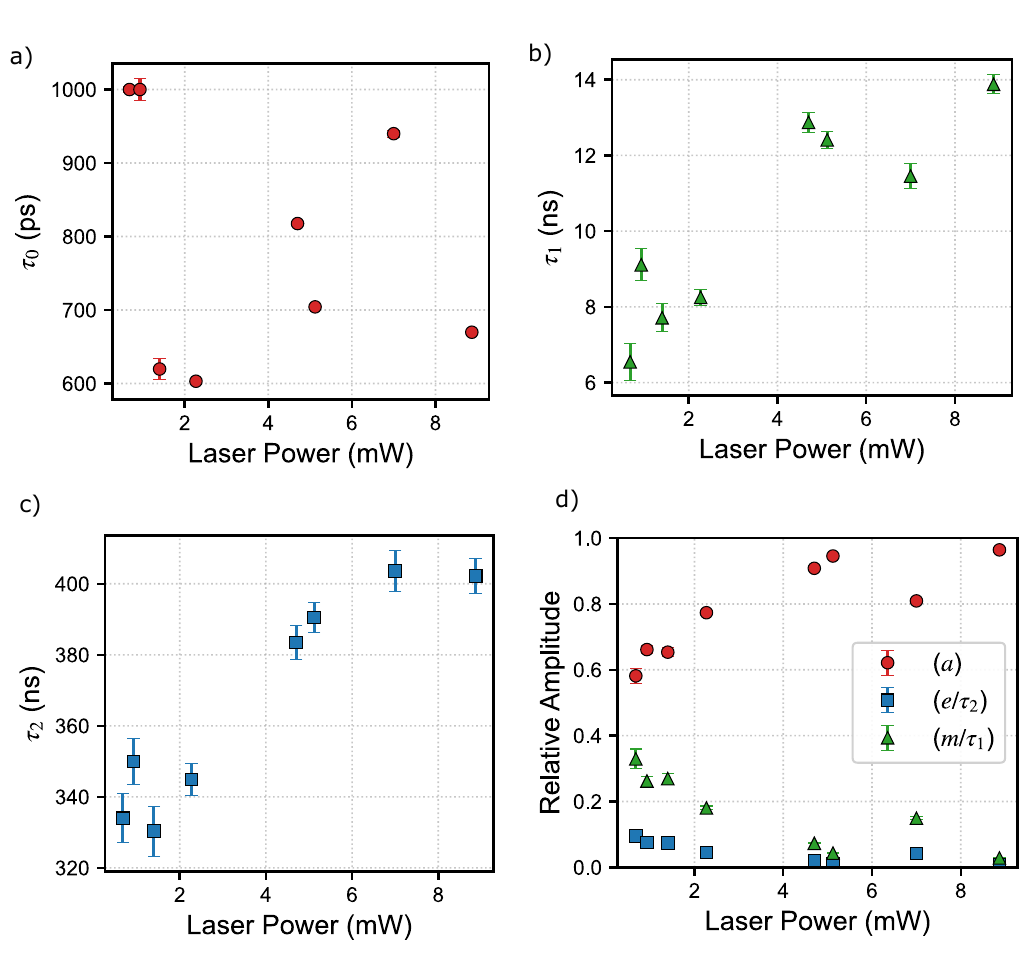}
    \caption{Excitation-fluence dependence of the defect-emission
    dynamics. (a) The fast component, $\tau_{0}$, does not exhibit a
    clear reproducible variation with excitation fluence and remains
    close to the temporal resolution of the instrument. The
    characteristic timescales (b) $\tau_{1}$ and (c) $\tau_{2}$  didnt show
     decrease with
    increasing fluence and rather fluctuates. (d) Relative contributions of the three decay
    channels. The contribution of the fast exponential component
    increases with excitation fluence, whereas the relative
    contributions of the two power-law components decrease.}
    \label{laserpower}
\end{figure*}

Excitation-fluence-dependent time-resolved PL measurements were
performed to examine whether exciton--exciton annihilation (EEA) makes
a significant contribution to the measured defect-emission dynamics.
EEA is an Auger-like many-body process in which the non-radiative
recombination of one exciton transfers its energy to a second exciton.
The exciton population in the presence of EEA may be expressed as

\begin{equation}
\frac{dn}{dt}
=
-\frac{n}{\tau}
-
\gamma_{\mathrm{EEA}}n^{2},
\end{equation}

where $\tau$ represents the lifetime in the low-density limit and
$\gamma_{\mathrm{EEA}}$ is the annihilation coefficient. Because the
second term becomes increasingly important at high exciton density,
EEA generally produces an excitation-density-dependent acceleration
of the early-time decay and a reduction in the effective exciton
lifetime\cite{Mouri_2014,erkensten2021dark}.

As shown in Fig.~\ref{laserpower}(a), the fitted value of $\tau_{0}$
does not exhibit a clear reproducible dependence on excitation
fluence. Since $\tau_{0}$ lies close to the instrument-response
function, small variations in this parameter cannot be interpreted
reliably. In contrast, neither $\tau_{1}$ nor $\tau_{2}$ exhibits the
systematic decrease expected if the measured long-time dynamics were
dominated by EEA. Both characteristic timescales which are fluctuating with fluence is shown in
[Fig.~\ref{laserpower}(b,c)]. It should be noted that laser power mentioned in x axis corresponds to peak laser power. Pulse period, pulse width and excitation spot size mentioned in previous section has to be used for calculating fluence.

It should be emphasized that $\tau_{1}$ and $\tau_{2}$ are the
characteristic parameters of the two power-law lifetime distributions,
rather than discrete single-exponential lifetimes. Their fluence
dependence can therefore also be influenced by trap filling and changes
in the relative populations of localized states. The absence
of fluence-induced shortening indicates that EEA is not the dominant mechanism
governing the observed long-lived defect-emission dynamics over the
investigated fluence range.

The relative contribution of the fast exponential channel increases
with excitation fluence, whereas the contributions of the two
power-law channels decrease [Fig.~\ref{laserpower}(d)]. This behavior of $\tau_{1}$ and $\tau_{2}$ 
is consistent with progressive filling or saturation of the finite
population of localized states, which reduces their relative
contribution as the excitation density increases. 

\subsection{Different fitting}
\begin{figure}[htbp]
    \centering
    \includegraphics[
        width=\textwidth,
        height=0.82\textheight,
        keepaspectratio
    ]
    {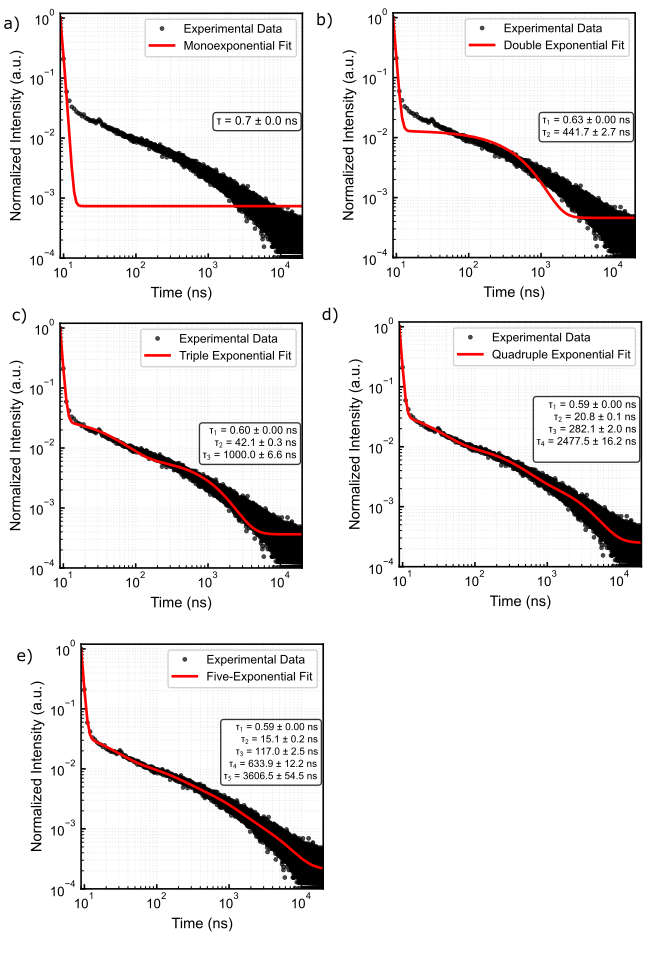}
    \caption{ Defect peak at 4K of WSe$_2$ fitted to different model a) Mono-exponential model b) Bi-exponential model c) Triple exponential model. d) Quadraple exponential model e) Five exponential model}
    \label{differentfit}
\end{figure}

The defect-related photoluminescence (PL) decay in WSe$_2$ was analyzed by systematically testing a hierarchy of kinetic models to understand the complex recombination dynamics. The fitting function for all models was of the form $I(t) = \text{model}(t) + \text{bkg}$, where  $\text{bkg}$ is a constant background.

We began with discrete exponential models, starting from a single component as shown in Fig \ref{differentfit} a:
\begin{equation}
I_1(t) = A e^{-t/\tau}
\end{equation}
which yielded $A = 0.998 \pm 0.001$, $\tau = 0.663$ ns.

Proceeding to multi-exponential models, we tested a bi-exponential function:
\begin{equation}
I_2(t) = A_1 e^{-t/\tau_1} + A_2 e^{-t/\tau_2}
\end{equation}
which revealed a fast component ($\tau_1 = {0.633}$ ns, 98.8\%) and a slow component ($\tau_2 = {441 \pm 2.696}$ ns, 1.2\%). As shown in Fig \ref{differentfit} b, fitting is poor.  
A tri-exponential model:
\begin{equation}
I_3(t) = A_1 e^{-t/\tau_1} + A_2 e^{-t/\tau_2} + A_3 e^{-t/\tau_3}
\end{equation}
further resolved the dynamics into fast ($\tau_1 = \SI{0.601}{\nano\second}$, 97.3\%), medium ($\tau_2 = \SI{42}{\nano\second}$, 2.1\%), and slow ($\tau_3 = \SI{1000}{\nano\second}$, 0.7\%) components. Again fitting is bad as shown in Fig \ref{differentfit} c.
Extending to a four-exponential model:
\begin{equation}
I_4(t) = A_1 e^{-t/\tau_1} + A_2 e^{-t/\tau_2} + A_3 e^{-t/\tau_3} + A_4 e^{-t/\tau_4}
\end{equation}
gives bad fit, yielding lifetimes of $\SI{0.5}{\nano\second}$ (96.7\%), $\SI{20.0}{\nano\second}$ (2.2\%), $\SI{282.0}{\nano\second}$ (0.8\%), and $\SI{2477}{\nano\second}$ (0.3\%). Fit data is shown in Fig \ref{differentfit} d. Thus, if we stick to conventional exponential function, minimum of five exponentials are required.

A satisfactory representation of the decay could eventually be obtained
using a five-exponential function,

\begin{equation}
I_5(t)
=
\sum_{i=1}^{5}
A_i\exp\left(-\frac{t}{\tau_i}\right)
+
\mathrm{bkg}.
\label{eq:five_exp}
\end{equation}

The fit yielded characteristic decay times of
$\tau_1 = 0.588 \pm 0.001$ ns,
$\tau_2 = 15.09 \pm 0.18$ ns,
$\tau_3 = 117.03 \pm 2.49$ ns,
$\tau_4 = 633.86 \pm 12.23$ ns, and
$\tau_5 = 3606.46 \pm 54.55$ ns.
The corresponding amplitude fractions were approximately
96.5\%, 2.0\%, 0.8\%, 0.5\%, and 0.2\%, respectively.
Thus, although the decay can be reproduced using a sufficiently large
number of exponential components, the resulting model requires ten
independent fitting parameters, comprising five amplitudes and five decay
times.

The mathematical success of the five-exponential model does not
necessarily imply the existence of five independent microscopic
recombination channels. A sum of discrete exponentials can approximate
a continuously distributed relaxation process when a sufficiently
large number of components is introduced. In such a fit, weak-amplitude
components may compensate for one another, and their extracted
timescales can depend on the initial parameter values, fitting bounds,
and temporal range of the measurement. Consequently, assigning a
distinct physical recombination mechanism to each of the five fitted
timescales would not be well justified.

Based on this physical insight, we developed a hybrid model that explicitly incorporates the power-law dynamics :
\begin{equation}
I_6(t) = a e^{-t/t_0} + d + \frac{m}{t + t_1} + \frac{e}{t + t_2}
\label{onebytequation}
\end{equation}
 This model captures the initial fast exponential decay. The inverse power-law terms $\frac{m}{t + t_1}$ and $\frac{e}{t + t_2}$ naturally describe the slow, non-exponential relaxation. This model has 4 less parameter than five exponential model . 
\section{DFT Calculations of Defect Configurations and Concentration Dependence}

\begin{figure}[htbp]
    \centering
    \includegraphics[
        width=\textwidth,
        height=0.82\textheight,
        keepaspectratio
    ]{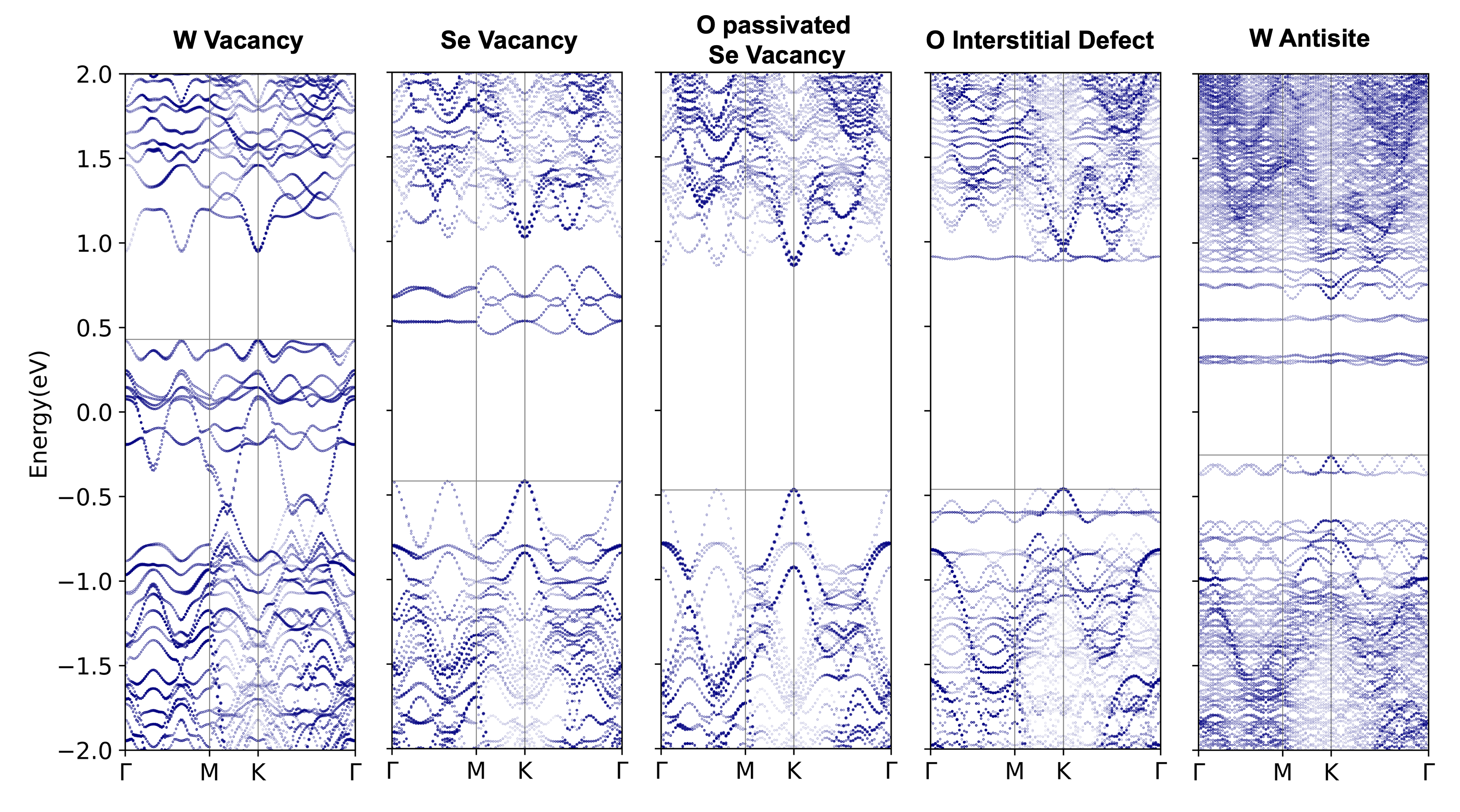}
    \caption{
    Calculated electronic band structures of monolayer WSe$_2$ containing
    different point-defect configurations: a W vacancy, a Se vacancy,
    substitutional O at a Se site, an O interstitial, and a W antisite.
    Substitutional O at a Se site does not produce prominent in-gap states,
    whereas the other investigated defect configurations introduce localized
    states within the band gap.
    }
    \label{def_iden}
\end{figure}

To examine the electronic states introduced by different point defects, spin-polarized density functional theory calculations were first performed using a $3\times3$ monolayer WSe$_2$ supercell. The defect configurations considered include a W vacancy, a Se vacancy, an O atom substituting a Se site, an O interstitial, and a W antisite defect. The calculated band structures are shown in Fig.~\ref{def_iden}. Among the investigated configurations, substitutional oxygen at a Se site does not introduce prominent electronic states within the band gap. In contrast, the W vacancy, Se vacancy, O interstitial, and W antisite configurations generate localized in-gap states with defect-dependent energies.

\begin{figure}[htbp]
    \centering
    \includegraphics[
        width=\textwidth,
        height=0.82\textheight,
        keepaspectratio
    ]{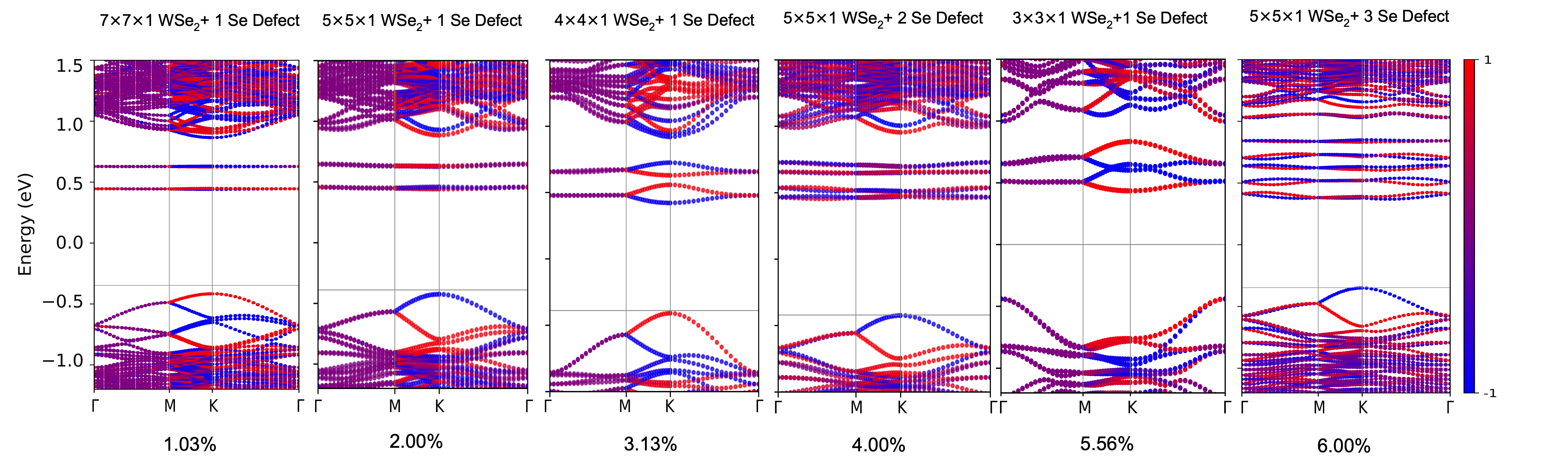}
    \caption{
    Dependence of the Se-vacancy electronic structure on defect concentration.
    The upper set of calculations corresponds to one Se vacancy introduced into
    WSe$_2$ supercells of different sizes, ranging from $3\times3$ to
    $7\times7$. Increasing the supercell size reduces interactions between
    periodically repeated defect images and leads to progressively flatter
    in-gap bands with weaker spin splitting. Additional calculations performed
    using a fixed $5\times5$ supercell with one, two, and three Se vacancies
    show that each vacancy introduces approximately two principal defect-state
    manifolds.
    }
    \label{def_con}
\end{figure}

The dependence of the Se-vacancy electronic structure on defect concentration was subsequently investigated by introducing one Se vacancy into progressively larger WSe$_2$ supercells, ranging from $3\times3$ to $7\times7$, as shown in Fig.~\ref{def_con}. Increasing the supercell size reduces the effective vacancy concentration and increases the separation between periodically repeated defect images. The calculated defect bands consequently become less dispersive, while the splitting between the spin-up and spin-down in-gap states decreases. This behaviour indicates that the dispersion observed in smaller supercells originates primarily from interactions between periodically repeated vacancies. The changes become substantially smaller beyond the $5\times5$ supercell, and the band structures obtained using the $5\times5$ and $7\times7$ cells are nearly flat. The defect bands in this dilute limit are therefore almost flat and exhibit only weak spin splitting.

The number of vacancies was also varied while keeping the supercell size fixed at $5\times5$ , as shown in Fig.~\ref{def_con}. The calculations show that an isolated Se vacancy produces two principal defect-state manifolds within the band gap. Accordingly, supercells containing two and three Se vacancies exhibit approximately four and six defect manifolds, respectively. These results indicate that the number of in-gap defect-state groups increases with the number of Se vacancies, while their detailed energies and dispersions are influenced by vacancy--vacancy interactions and the local atomic arrangement.

%

\end{document}